\documentclass[twocolumn,amssymb]{revtex4} 
\usepackage{epsfig,amssymb,amsmath,graphicx,subfigure,hyperref  }  
\usepackage{color}

\begin{document}

\title{Depletion zones and crystallography on pinched spheres} 
\author{Jingyuan Chen$^{1,2}$, Xiangjun Xing$^{1,3}$ and Zhenwei Yao$^{1,2}$} 
\email{zyao@sjtu.edu.cn}
\affiliation{$^1$School of Physics and Astronomy, and $^2$Institute of Natural Sciences, Shanghai Jiao
      Tong University, Shanghai 200240 China \\
$^3$Collaborative Innovation Center of Advanced Microstructures, Nanjing 210093, China}  
\begin{abstract} 
Understanding the interplay between ordered structures and substrate curvature
  is an interesting  problem with versatile applications, including
  functionalization of charged supramolecular surfaces and  modern microfluidic
  technologies. In this work, we investigate the two-dimensional packing
  structures of charged particles confined on a pinched sphere. By continuously
  pinching the sphere, we observe cleavage of elongated scars into pleats,
  proliferation of disclinations, and subsequently,  emergence of a depletion
  zone at the negatively curved waist that is completely void of particles. We
  systematically study the geometrics  and energetics of the depletion zone, and
  reveal its physical origin as a finite size effect, due to the interplay
  between Coulomb repulsion and concave geometry of pinched sphere. These results further our understanding of
  crystallography on curved surfaces, and have implications in design and
  manipulation of charged, deformable interfaces in various applications. 
\end{abstract}
\maketitle

\section{Introduction}

Electrically charged and deformable interfaces are widely seen in natural and synthetic two-dimensional systems, ranging from charged droplets of spherical~\cite{rayleigh1882xx, duft2002shape, liu2015dynamics}, 
cylindrical~\cite{huebner1971instability, guerrero2014whipping} and toroidal~\cite{pairam2009generation,
fragkopoulos2017toroidal} geometries, various charged
elastic ribbons fabricated by supramolecular
self-assembly~\cite{aggeli2001hierarchical, weingarten2014self} to charged lipid
membranes~\cite{andelman1995electrostatic, lipowsky1995structure,
  genet2001influence}.  Surface charges
  can strongly affect morphologies of deformable interfaces~\cite{kim2003effects,vernizzi2007faceting,
  adamcik2010understanding,  
  yao2016electrostatics}, renormalize their elastic
rigidities~\cite{winterhalter1988effect,lau1998charge,nguyen1999negative,netz2001buckling},
and create local electrostatic structures with chemical and biological
implications~\cite{genet2001influence,
weingarten2014self}.  Proper understanding
of the subtle interplay between long-range Coulomb interaction and surface
morphology is not only interesting from a theoretical point of view, but also a prerequisite for applications of charged deformable interfaces, including morphology-based functionalization of charged supramolecular surfaces~\cite{palmer2007supramolecular, moyer2012tuning, weingarten2014self} and modern microfluidic technologies involving electrostatic dispersion of liquids~\cite{taylor1964disintegration,ziabicki1976physical,bailey1988electrostatic,gomez1994charge,Urbanski2017}. As a classic example,  Lord Rayleigh demonstrated that  Coulombic repulsion drives large deformation and fission of a charged liquid droplet  long ago~\cite{rayleigh1882xx}. 

Another interesting example is the Thomson problem~\cite{thomson1904xxiv,nelson2002defects,de2005electromagnetic,bowick2009two,koning2014crystals}, which is to find the ground state of  charged particles confined on a spherical surface. In
the resulting triangular lattice on the sphere, several curvature-driven
crystallographic defect motifs like scars~\cite{bausch2003grain}
and vacancies~\cite{meng2014elastic,yao2017topological} have been discovered in the extensive
theoretical~\cite{altschuler1997possible,bowick2002crystalline,vitelli2006crystallography,burke2015role, mehta2016kinetic,yao2016dressed}
and experimental~\cite{bausch2003grain,irvine2010pleats,irvine2012fractionalization,meng2014elastic} studies. By allowing the surface to be deformable, the generalized Thomson problem can be employed to explore the coupling of charged particles distribution and the deformation of the interface.  This can be experimentally realized by depositing electrically charged colloids at the spherical water-oil
interface~\cite{bausch2003grain,irvine2010pleats}, or by encapsulating an aqueous
mixture of colloids of different sizes within spherical water-in-oil
droplets~\cite{meng2014elastic}.

In this work, we shall control the deformation of interface, and study the
lowest-energy distribution of the charges that are mobile on the surface.
Specifically, we pinch a sphere around the equator to form a peanut-like shape.  The pinching creates  a negatively curved region near the waist. Our study shows that this has strong influence on the packing structure of charges, leading to formation of the elongated scars which further evolve into pleats, as well as  formation of cracks that ultimately evolve to a depletion zone, where all charges are excluded. 

We systematically study the geometric structure of the depletion zone, and
demonstrate that it is a robust feature of the lowest-energy states, resulting
from the interplay between Coulombic repulsion and the concave geometry of the
waist.  Furthermore, heuristic arguments in terms of continuum electrostatics show that the depletion zone structure is a finite size effect. Finally, we briefly discuss packing of charges on a biconcave discoidal shape, and find similar defect patterns and depletion zones. The discovery of depletion zone structure over the concave geometry enriches our understanding of the crystallographic defects in curved crystals, and may have useful applications in the design and manipulation of charged deformable interfaces.

\section{Model and Method}

\begin{figure*}[htp!]  
\centering 
\subfigure[]{
\includegraphics[width=1.25in]{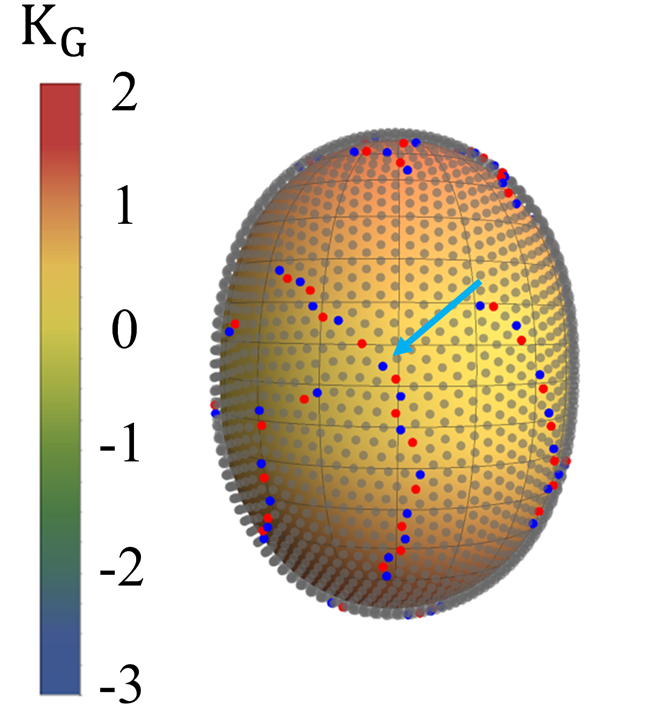}}
\hspace{-0.1in}
\subfigure[]{
\includegraphics[width=1.25in]{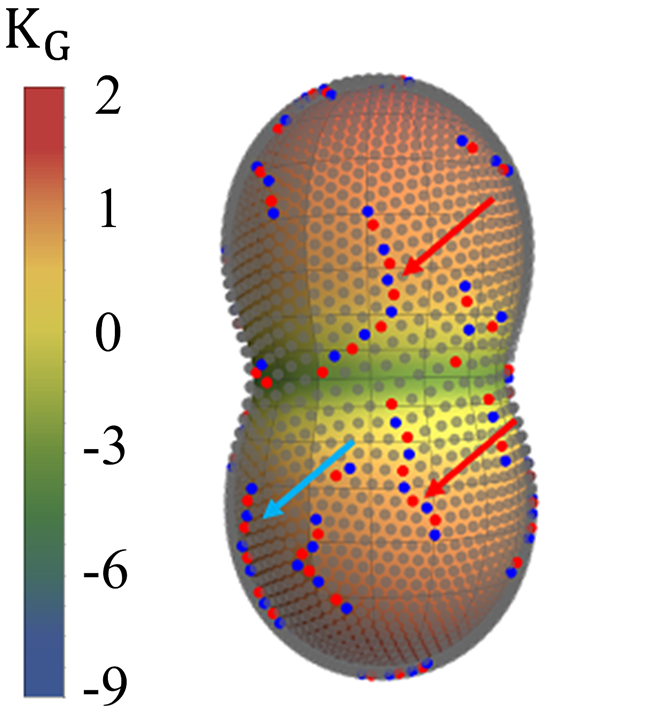}}
\hspace{-0.1in}
\subfigure[]{
\includegraphics[width=1.25in]{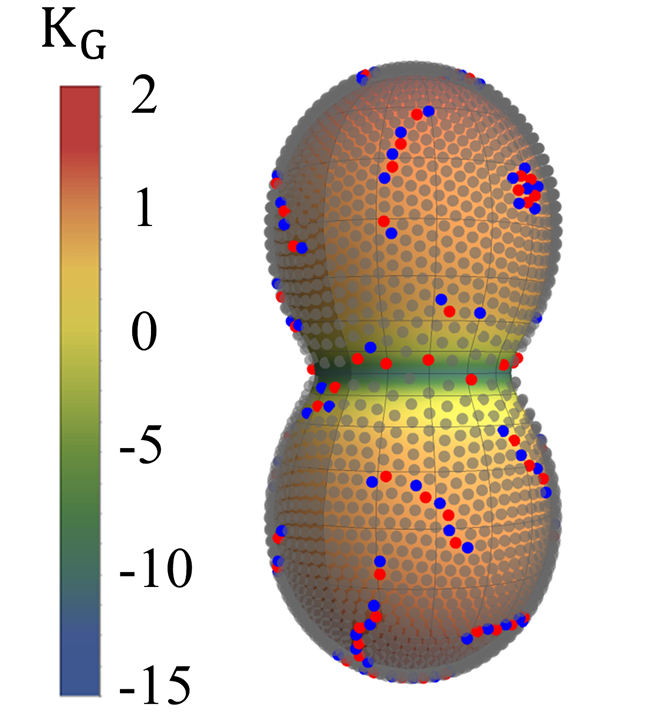}}
\hspace{-0.1in}
\subfigure[]{
\includegraphics[width=1.25in]{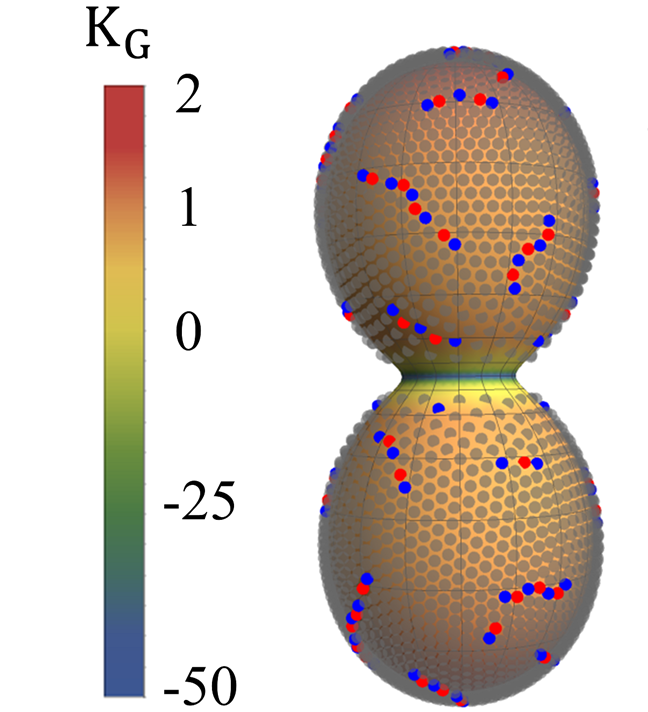}}
\hspace{-0.1in}
\subfigure[]{
  \includegraphics[width=1.8in]{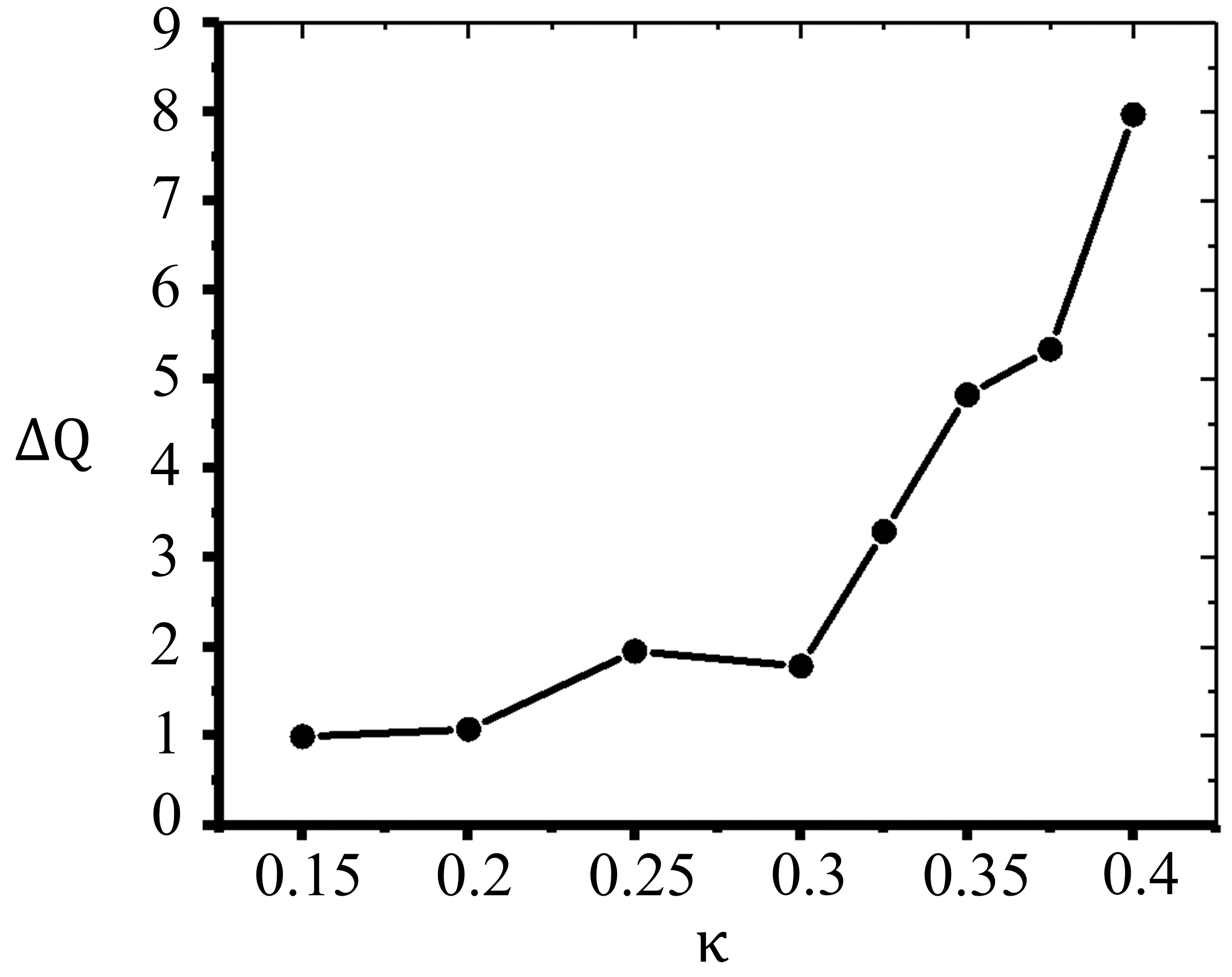}}
\caption{Lowest-energy particle configurations over the typical pinched spheres. The
  degree of the pinching deformation is characterized by $\kappa$. (a) $\kappa=0.1$, (b)
    $0.3$, (c) $0.4$, and (d) $0.6$. The five- and
    seven-fold disclinations are colored in red and green, respectively. The
    rightmost panel shows the distribution of the mean curvature. (e) Plot of the excess topological charge $\Delta Q$ in the waist region over the pinched sphere versus $\kappa$. $N=3000$. }
\label{fig1}
\vspace{-2mm}
\end{figure*}

A pinched sphere can be described in terms of the spherical harmonics as 
\begin{align} r(\theta)&= R \left[ 1+\kappa (3 \cos^2\theta  -1 )\right], 
  \label{shape2}
\end{align}
where $\theta$ is the polar angle, and  $\kappa$  is a dimensionless parameter controlling the degree of pinching.   For $0<\kappa<1$, the pinched sphere exhibits a peanut-like shape, as illustrated in Fig.~\ref{fig1}. The radius achieves minimum at the {\em waist}, with $r(\theta=\pi/2)=(1-\kappa)R$, which shrinks as $\kappa$ increases.   For $-1/2< \kappa<0$, the pinched sphere  becomes a biconcave discoid, illustrated in Fig.~\ref{fig6}.

The Gaussian curvature~\cite{struik88a} of a surface is given by
\begin{align}
K_{\textrm G}=  {1}/{R_1R_2}, 
  \end{align}
where $R_1$ and $R_2$ are two radii of  principal curvatures.  For a pinched sphere Eq.~(\ref{shape2}), the Gaussian curvature is
\begin{align}
K_{\textrm G}=-\frac{1}{2\sqrt{|g|}}\frac{\partial}{\partial
    \theta}\left(\frac{1}{\sqrt{|g|}}
    \frac{\partial}{\partial \theta} 
  (r^2 \sin ^2\theta)
    \right).
\end{align}
where  $|g| $ is the determinant of the metric tensor.  For $\kappa>0$, $K_{\textrm G}$ reaches minimum at the waist circle $\theta = \pi/2$:
\begin{eqnarray}
K_{\textrm {G,min} } = \frac{1-7\kappa}{(1-\kappa)^3}.
\end {eqnarray}
$K_{\textrm {G,min} }$ becomes zero when  $\kappa={1}/{7}$, and diverges as
  $\kappa \rightarrow 1$.  Hence, for ${1}/{7} < \kappa <1$, the waist area is negatively curved. According to the Gauss-Bonnet-Chern theorem\cite{greub1973lie}, the integral of the Gaussian curvature over a smooth surface of spherical topology is $4\pi$, which is a topological invariant.

We perform simulations to find the lowest-energy state of a collection of point
  charges confined on the pinched sphere Eq.~(\ref{shape2}). These charges
  interact with each other via the three-dimensional Coulomb potential $V_{ij}
  \sim 1/r_{ij}$, where $r_{ij}$ is the distance between particle $i$ and $j$.
  We find the lowest-energy state by starting from a random initial state
  (unless specified otherwise) and let the particles move incrementally along
  the forces acting on them.   We update the particle configuration in a
  collective manner to rapidly reduce the system energy, followed by moving
  individual particles in each simulation step to fine tune the system down to
  the bottom of the energy surface. This protocol has been applied in several
  charged particle systems and has successfully generated lowest-energy
  states~\cite{yao2013topological, yao2016electrostatics}.

\begin{figure}[th]
\centering 
\subfigure[]{
	\includegraphics[width=1.6in]{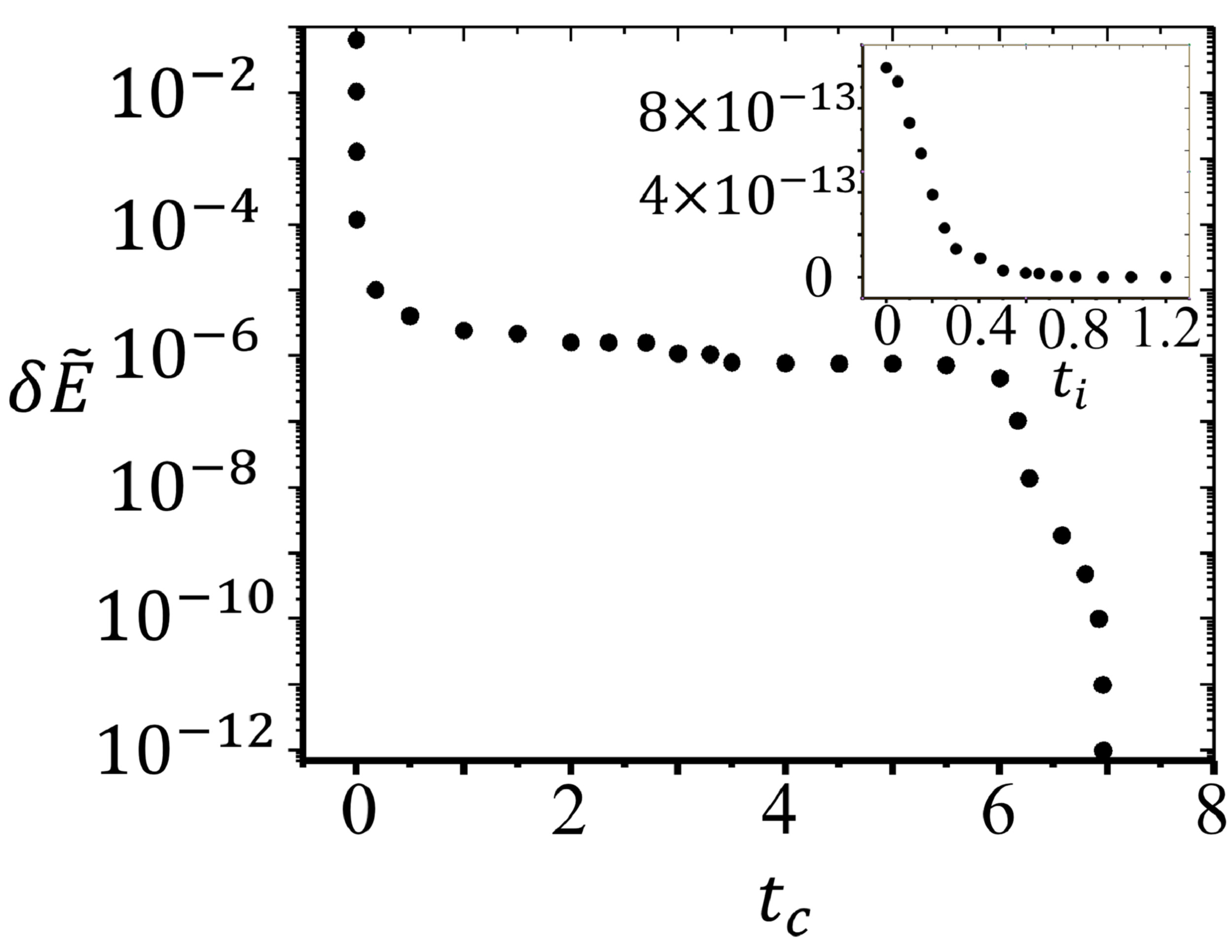}}
\subfigure[]{
	\includegraphics[width=1.62in]{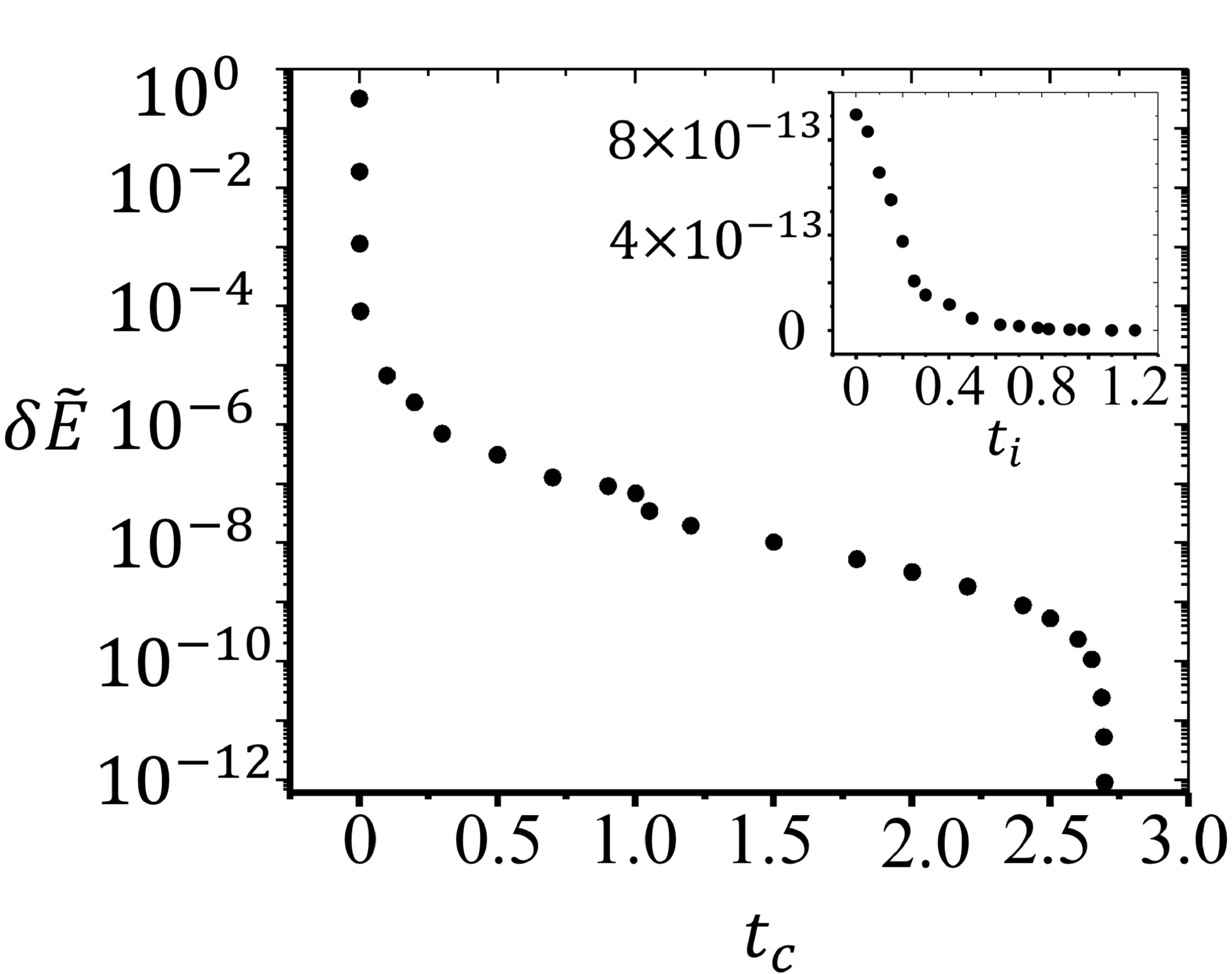}}
\vspace{-3mm}
  \caption{Convergence of system energy towards the lowest value for the cases
  of (a) $\kappa= 0.15$ and (b) $\kappa=0.71$. $\delta\tilde{E}=(E_e-E_f)/E_f$,
  where $E_e$ is the total electrostatic energy and $E_f$ is that in the lowest
  energy state. In the collective movement of particles, step size is gradually
  reduced to accelerate energy reduction.   The unit of $t_c$ is 10000 sweeps
  over all the particles, and that of $t_i$ in the individual mode (shown in the
  insets) is by sweeping 10000 particles. $N=3000$. }
\vspace{-2mm}
\label{fige}
\end{figure}

  In Fig.~\ref{fige}(a) and \ref{fige}(b), we show the reduction of system
  energy towards the lowest value for the cases of $\kappa= 0.15$ and $\kappa=0.71$,
  respectively. $\delta\tilde{E}=(E_e-E_f)/E_f$, where $E_e$ is the total
  electrostatic energy and $E_f$ is that in the lowest energy
  state. The unit of $t_c$ is 10000 sweeps over all the particles. In the
  collective movement of particles, we gradually reduce the step size from
  $2a_0$ to $7.8\times10^{-4}a_0$ for efficiency in energy reduction. $a_0$ is
  the lattice spacing. We further reduce system energy by moving particles
  individually, and the results are presented in the insets in Fig.~\ref{fige}.
  We see that the system energy rapidly converges after a few sweeps. Note that
  the unit of $t_i$ in the individual mode is by sweeping 10000 particles.

We analyze the crystallographic defects in the resulting lowest-energy states
  using the method of Delaunay triangulation~\cite{nelson2002defects}.  {While
  this is a standard procedure of triangulation on  plane, the existence and
  uniqueness of triangulated state using this method on a negatively curved
  surface is not guaranteed.  To solve this problem, we choose a disk-shape
neighborhood of each particle, and project it  on to the tangent plane.  As long
as the disk radius is sufficiently small, the curvature within the disk can be
ignored.  By performing the standard Delaunay triangulation over these particles
projected to the plane, we can determine the coordination number of each
particle without  unambiguity.

To check the validity of our simulation codes, we  first simulate the evolution
  of charged particles on an unperturbed spherical substrate.  Starting from an
  initial state with dense distribution of disclinations, we numerically observe
  the annihilation of disclinations.  Only a few scattered scars and
  dislocations survive in the final lowest-energy states that our simulations
  can reach.  These observations are consistent with the known results in
  spherical crystals, suggesting the reliability of our simulations.

\begin{figure*}[ht!]  
\centering 
\subfigure[]{
\includegraphics[width=1.65in]{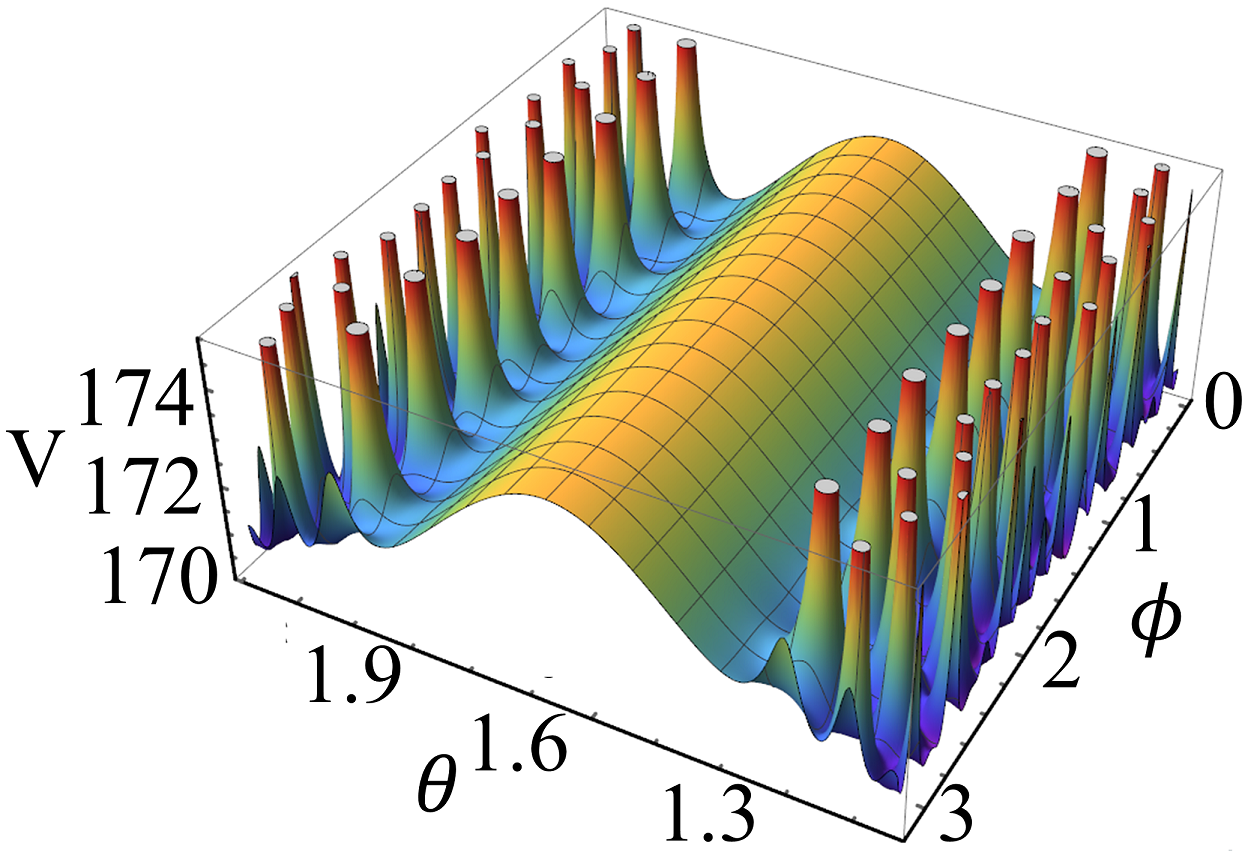}}
\subfigure[]{
\includegraphics[width=1.65in]{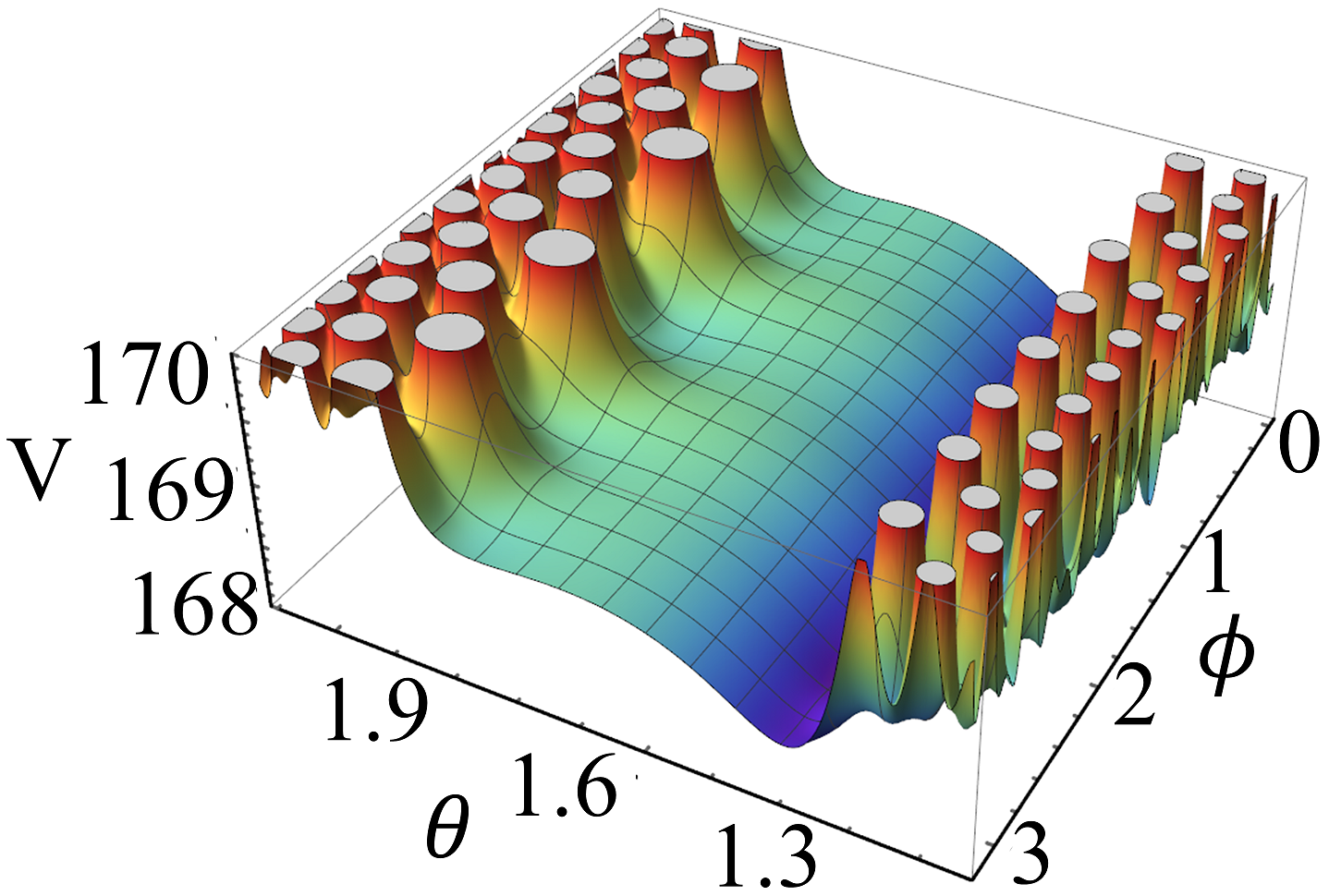}}
\subfigure[]{
\vspace{-3mm}
\includegraphics[width=1.67in]{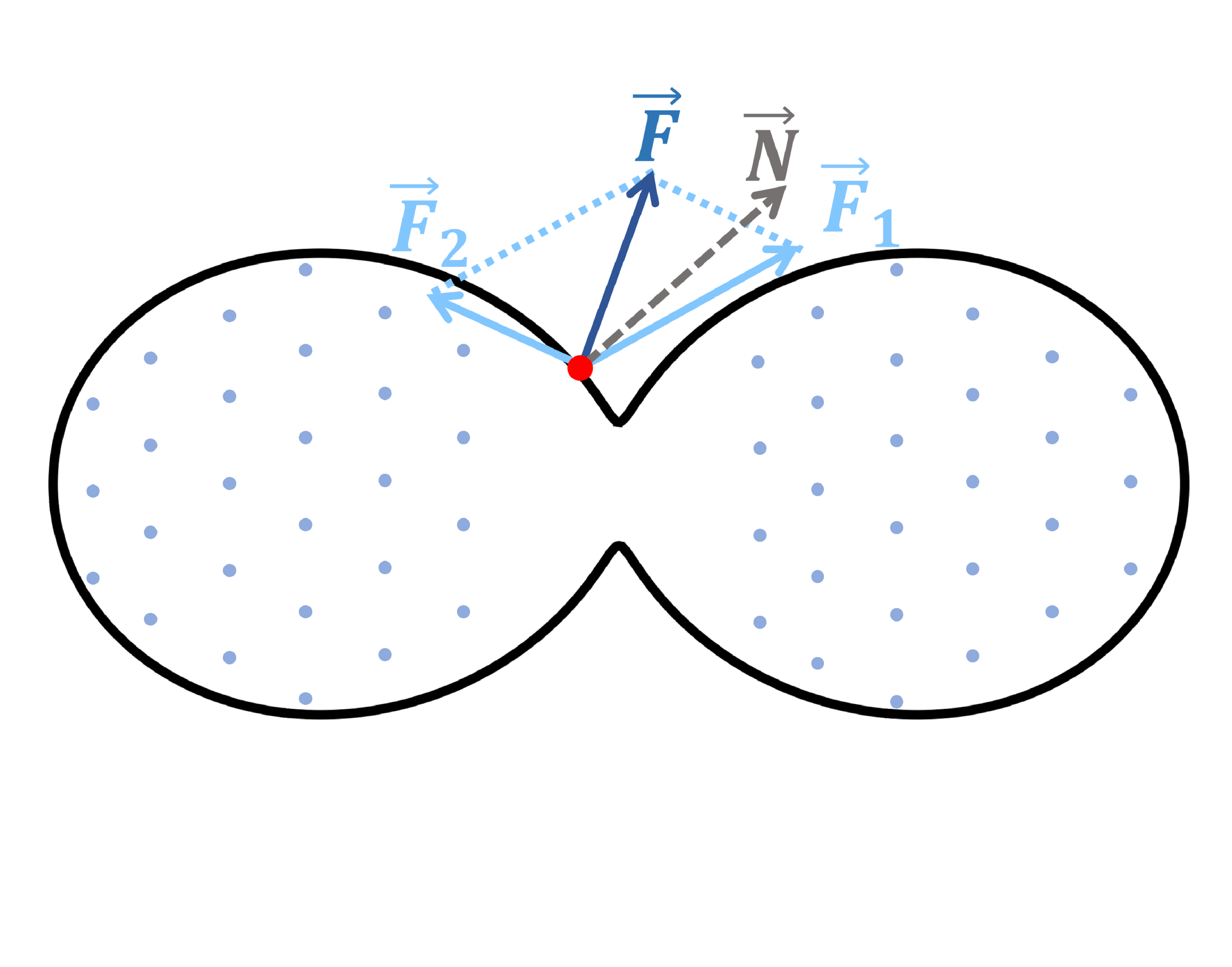}}
\caption{The electric potential near the waist.  $\theta, \phi$ are polar and azimuthal angles, respectively. (a) Equal number of particles in two sides of the waist.  There is a potential maximum on the waist.  Hence a test charge cannot be stabilized near the waist.  (b) Particle number in the two sides are slightly different.  There is no potential barrier near the waist, and a test charge will automatically move to the right side with fewer particles. $\kappa=0.7$. (c) Schematic plot of the force analysis on a test charge indicated by the red dot in the
    depletion zone (see text for more information). 
}
\vspace{-2mm}
\label{fig2}
\end{figure*}

\section{Results and Discussion}

\subsection{From elongated scars to pleats}

Past experimental~\cite{bausch2003grain} and theoretical~\cite{bowick2002crystalline} studies have   established that the charges confined on a sphere will spontaneously organize into a   triangular lattice, with various sorts of defects~\cite{nelson2002defects,bowick2009two,koning2014crystals}, including disclinations, dislocations and scars.  A z-fold vertex, i.e. a particle with coordination number $z$, is called a disclination if $z$ differs from six.  The  topological charge of a disinclination is defined as $q = (6- z )\pi/3$.
  According to  Euler's theorem~\cite{struik88a}, the total topological charge of a spherical crystal is a topological invariant:  $\sum_{i}q_i = 4\pi$, where the sum is over all particles  of lattice.  A dislocation is a pair of five- and seven-fold disclinations, with vanishing net topological charge.  A scar is a string of alternating five- and seven-fold disclinations with two five-fold disclinations at the ends.  The topological charge of a scar is hence $\pi/3$, the same as a single five-fold disclination.  A scar is essentially a grain boundary across  which a mismatch of crystallographic orientation occurs. 

It has been established by previous studies that in a sufficiently large
  spherical crystal, the isolated point-like five-fold disclinations become
  unstable and crack into scars~\cite{bausch2003grain}.    The length of the scars is determined by the ratio $R/a$, where $R$ is the spherical radius, while $a$ is the lattice spacing.   By slightly pinching the sphere along its equator, we observe an appreciable change of scar lengths.  Elongated scars are found over the pinched spherical crystal, as indicated by the blue arrow in Fig.~\ref{fig1}(a). These scars tend to align along the long axis (defined as the z-axis) of the pinched sphere. 

When $\kappa$ increases to about $0.3$, we find strings of alternating
disclinations with a five-fold disclination at one end and a seven-fold
disinclination at the other end, as indicated by the red arrows in 
Fig.~\ref{fig1}(b).   Such defect strings are called {\it pleats}. Their net
topological charge is zero~\cite{irvine2010pleats}.  Pleats have been
first observed in the crystalline order on negatively curved capillary
bridges~\cite{irvine2010pleats}.  We notice that the negatively charged ends of the pleats,  indicated by the red arrows in Fig.~\ref{fig1}(b),  are anchored at the
negatively curved waist, while the positively charged end are located in the
positively curved region.  Tracking the evolution of  defect patterns as we tune
the pinching, we find that these two pleats were split out from the same
elongated scar. As a consequence of this splitting, the scar becomes significantly shorter.    
In this process, an extra short scar is emitted to the surrounding crystalline region to conserve the total topological charge.

Past studies have demonstrated that Gaussian curvature can be understood as
  uniform counter-charge background that screens the topological charges of
  disclinations~\cite{nelson2002defects}.  From this perspective, it is useful
  to define {\em the geometric charge} $Q_g$ of the waist area as the integral
  of the Gaussian curvature,  the {\em  excess charge} $\Delta Q$ as $\Delta Q =
  Q_d - Q_g$, where $Q_d$ is the total disclination charge.
  Figure~\ref{fig1}(e) shows that the excess charge $\Delta Q$ is always
  positive over the range of $\kappa \in [0.15, 0.4]$, indicating that geometric
  curvature can only partially screen the topological defects.  Furthermore,
  from Fig.~\ref{fig1}(e),  we see that $\Delta Q$ surges  abruptly as $\kappa$
  increases from 0.3 to 0.4, which coincides with the emergence of pleats and
  complete disruption of crystalline order near the waist, as shown in
  Figs.~\ref{fig1}(b) and \ref{fig1}(c).  Previous studies have shown that
  long-range repulsion driven particle-density gradient will induce a negative
  Gaussian curvature and promote emergence of seven-fold
  disclinations~\cite{nelson2002defects, Mughal2007}. This
  effect may partially explain the excess topological charges in the waist
  area.

 \vspace{-2mm}
\subsection{Depletion zone structure}

Further increasing $\kappa$ to $0.4$, we observe filling of seven-fold
 disclinations and complete disruption of the crystalline order near the waist,  as illustrated in Fig.~\ref{fig1}(c). Concomitantly, the particle density near the waist also becomes much lower: this seems a favorable strategy to lower the electrostatic
 energy of the system.  For $\kappa \geq 0.6$, we find the emergence of a sharply defined {\em depletion zone} (with no particle) around the waist, as shown in Fig.~\ref{fig1}(d).

We shall verify that the depletion zone is a robust feature of the ground state.
To this end, we start from initial states where only half of the pinched sphere
($\kappa=0.7$) is occupied by randomly distributed particles.  Once the dynamics
is turned on, we find particles migrate to the other side through the waist.  In
the final lowest-energy state, a depletion zone is formed again over the waist
region.  The particle numbers at  two sides of the waist differ by less than
$1\%$.  This indicates that the depletion zone is indeed a robust feature of
the ground sate and is independent of the initial condition.

\begin{figure}[th]  
\centering 
\subfigure[]{
\includegraphics[width=1.6in]{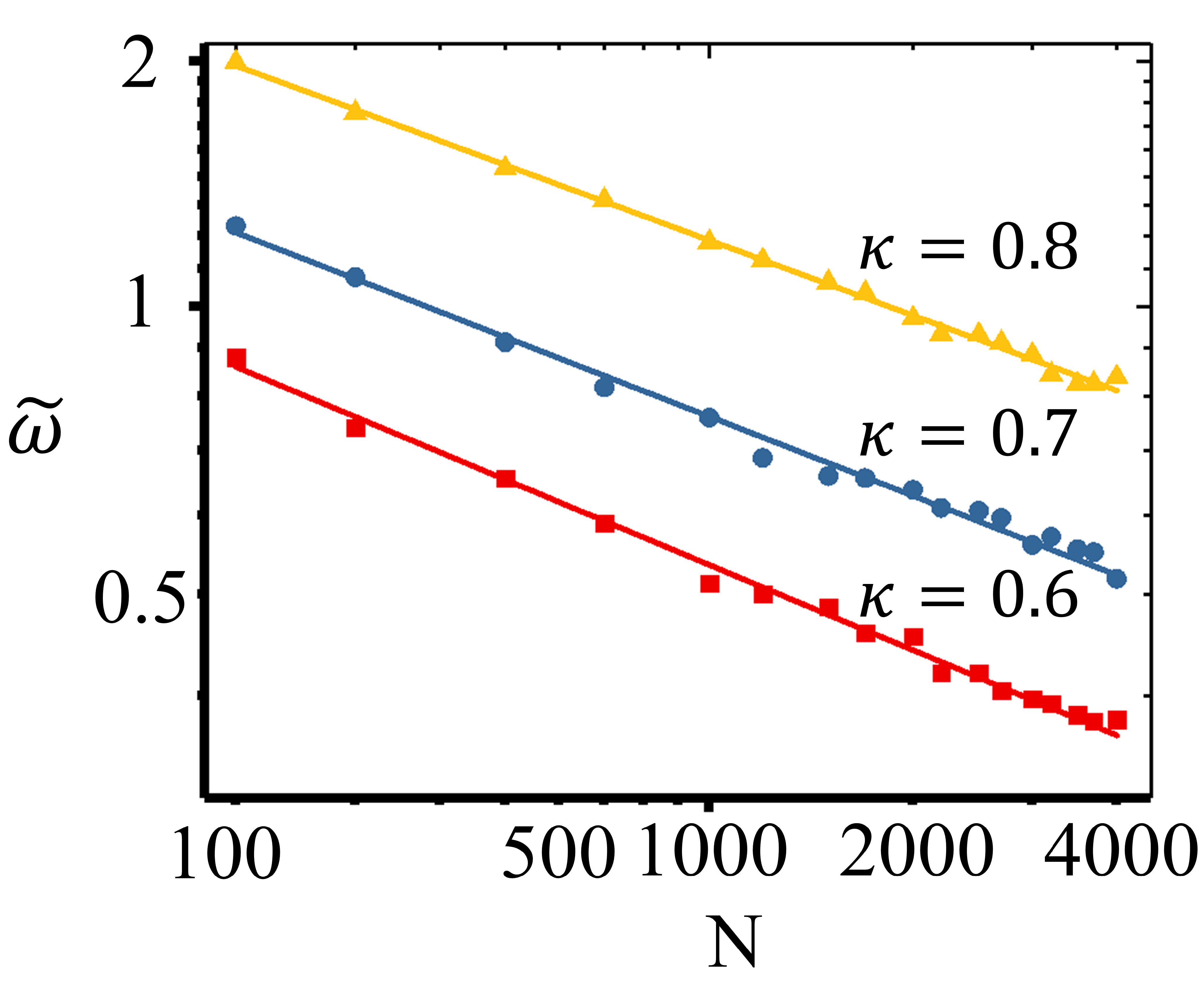}}
\subfigure[]{
\includegraphics[width=1.62in]{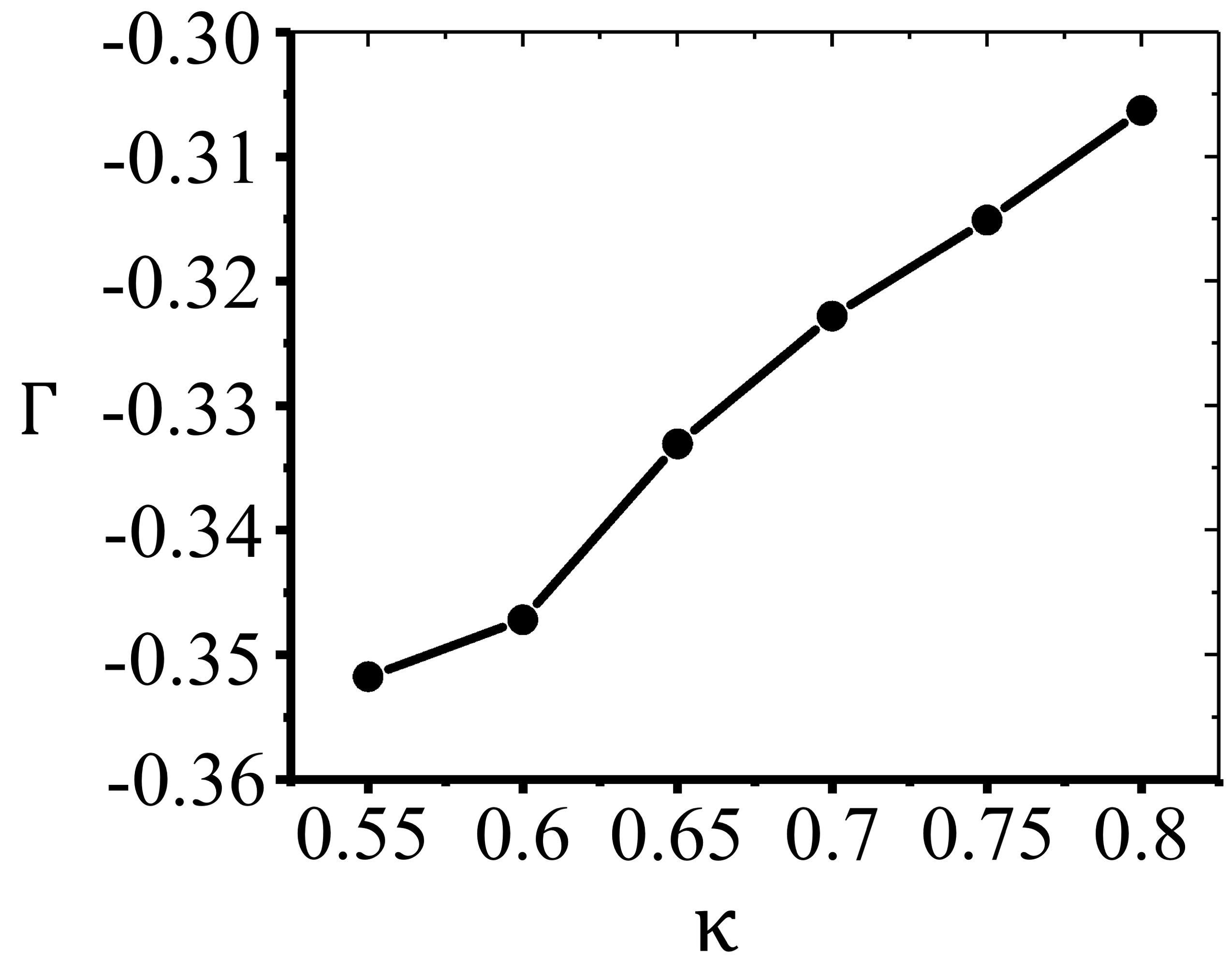}}
\vspace{-3mm}
\caption{Geometric characterization of the depletion zone structure in the
  lowest-energy pinched spherical crystals. (a) Logarithmic plot of the
  normalized width $\tilde{w}$ of the depletion zone versus the number of
  particles $N$. The data can be well fitted by a power law with the exponent
  $\Gamma$. (b) Plot of $\Gamma$ vs $\kappa$. $N=3000$.
  }
  \vspace{-2mm}
\label{fig3}
\end{figure}

We can also plot the electrostatic potential on the surface as a function of
polar and azimuthal angle $\theta, \phi$.  Figure \ref{fig2}(a) shows the case  where particle numbers in the two sides of the waist are equal.  It clearly shows that the potential has a maximum on the waist where $\theta = \pi/2$.  A  test charge (recall that all particles are positively charged) cannot be stabilized at the waist: it will move to either side.  By contrast,
Fig.~\ref{fig2}(b) shows the case where particle numbers in two sides are
slightly different (992 vs 1008 particles), and there is no potential
maximum at the waist.  The potential increases monotonically with the polar angle $\theta$.  A test charge placed near the waist will move to
the side with lower potential. These analyses further substantiate that  the depletion zone is indeed a robust property of the ground state. 

To better understand the potential maximum on the waist as shown in
Fig.~\ref{fig2}(a), let us consider the electrostatic force acting on a test
charge slightly to the left of the waist, as schematically shown in
Fig.~\ref{fig2}(c).  Since the test charge is confined on the surface, only the
tangent component of the total force is relevant.  Now, as the parameter $\kappa$
increases, the waist becomes thinner, and the local normal vector  $\vec{N}$
rotates to the right.  For sufficiently thin waist, $\vec{N}$ is always to the
right of $\vec{F}$, and hence the tangent component of $\vec{F}$ always points
to the left.  That is, the test charge is pushed to the left.  This is exactly
what we see in Fig.~\ref{fig2}(a).  The depletion zone is caused by the local
concave geometry near the waist.

We proceed to discuss the influence of particle number $N$ and pinching
parameter $\kappa$ on the size of the depletion zone. Let $\widetilde{\omega}$
be the width of depletion zone normalized by the diameter of waist circle.
Figure~\ref{fig3}(a) shows that $\widetilde{\omega}$ shrinks as $N$ increases,
with the dependence well-described by a power law for $N$ from 100 to 4000:
\begin{eqnarray} \widetilde{\omega} \sim N^{\Gamma}.\label{omega} \end{eqnarray}
  where the exponent $\Gamma $ increases linearly with $\kappa$, as shown in
  Fig.~\ref{fig3}(b). This suggests that the depletion zone is a finite size
  effect, and will eventually disappear in the limit  $N\rightarrow \infty$.
  Indeed, in this limit, we expect that the electrostatic potential is described
  by Laplace equation with {\em conductor boundary condition}, i.e., the
  potential is constant on the entire surface, since all charges are mobile in
  our model.  Now if the charge density vanishes in certain region on the
  surface, then both the potential and its normal derivative are fixed. This
  corresponds to Cauchy boundary conditions, which is known to be incompatible
  with Laplace equation.  We further note that since $|\Gamma| < 1/2$,
  $\widetilde{\omega}$ decays with $N$ slower than the mean inter-particle
  distance $\langle a \rangle$, which scales as $1/\sqrt{N}$.

\subsection{Negatively pinched sphere}

It is also interesting to study the case of negative $\kappa$, where the surface
has the shape of biconcave discoid.  The defect structures for several negative
values of $\kappa$ are presented in Fig.~\ref{fig6}.  At $\kappa=-0.1$, we find
scars [indicated by the blue arrow in Fig.~\ref{fig6}(a)] around the rim of the
pinched sphere where the Gaussian curvature reaches maximum, and pleats
(indicated by the red arrow) over the slightly curved areas. At $\kappa=-0.2$,
we find isolated seven-fold disclination at the dimple as shown in the inset of
Fig.~\ref{fig6}(b).  Depletion zone starts to appear when $\kappa$ is even more
negative as shown Figs.~\ref{fig6}(c) and \ref{fig6}(d) for the cases of
$\kappa=-0.3$, and $-0.4$.  As in the case of positive $\kappa$, these depletion
zones are expected to vanish in the limit $N \rightarrow \infty$.

\begin{figure}[t]  
\centering 
\subfigure[]{
\includegraphics[width=1.5in]{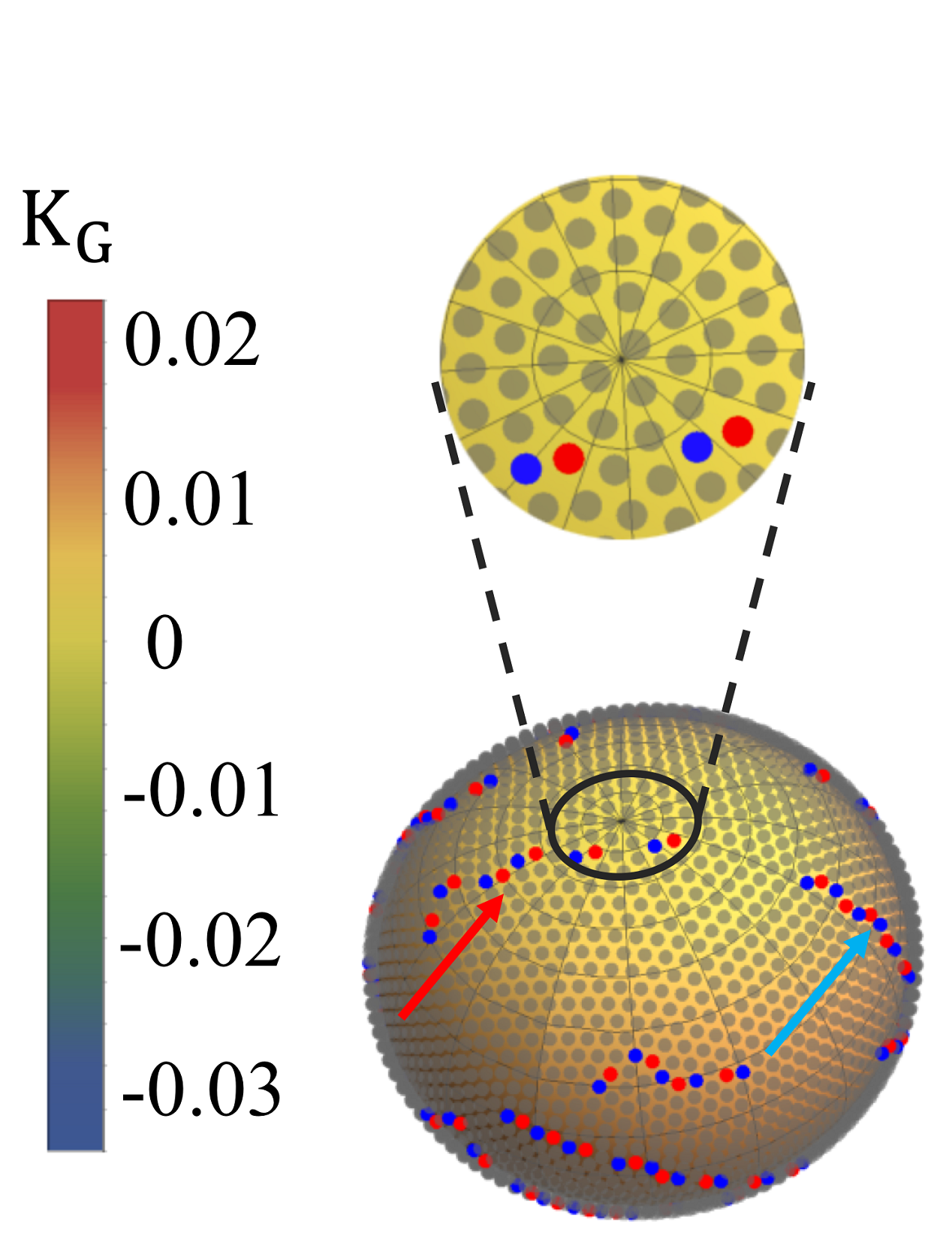}}
\hspace{-0.1in}
\subfigure[]{
\includegraphics[width=1.5in]{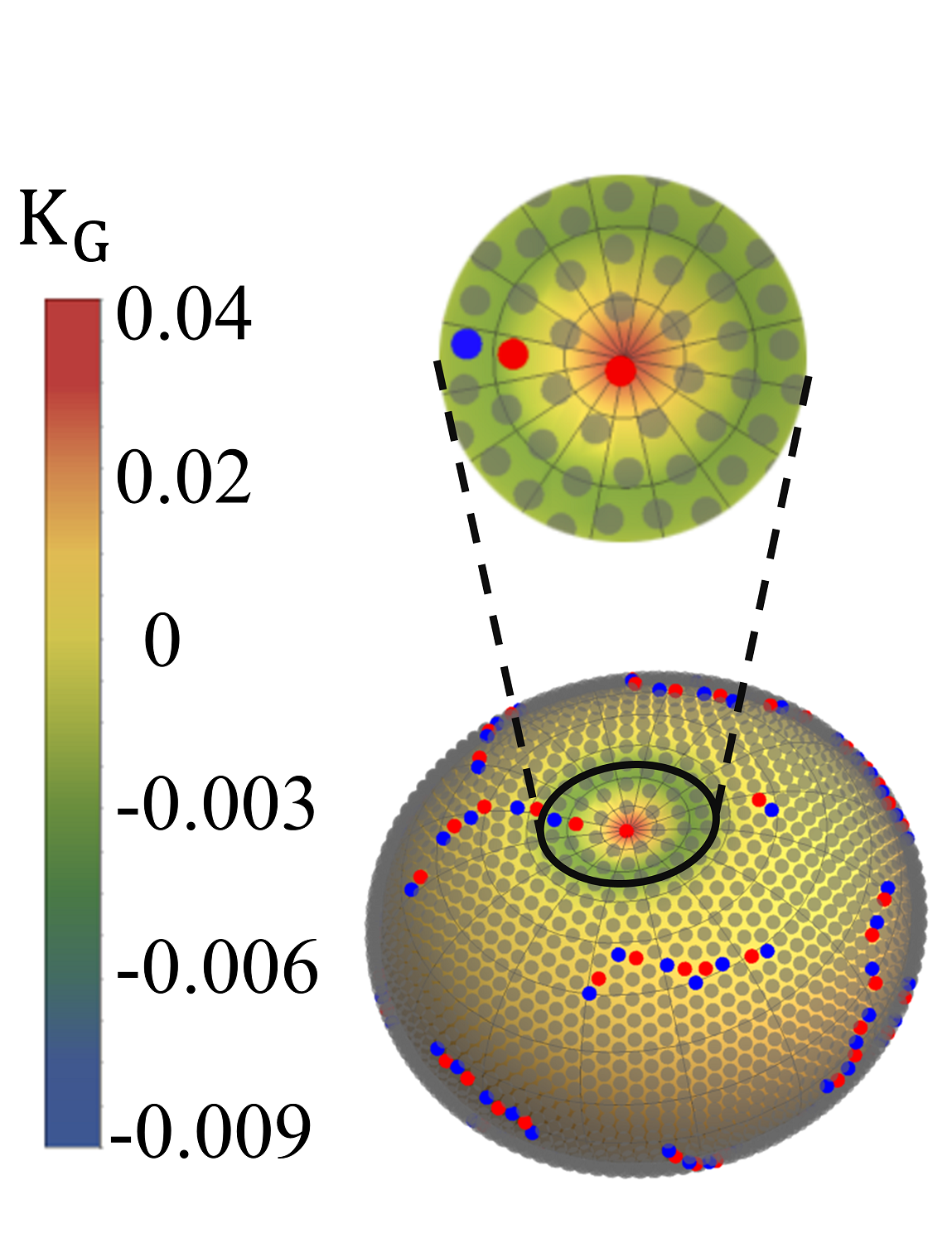}}
\subfigure[]{
\includegraphics[width=1.5in]{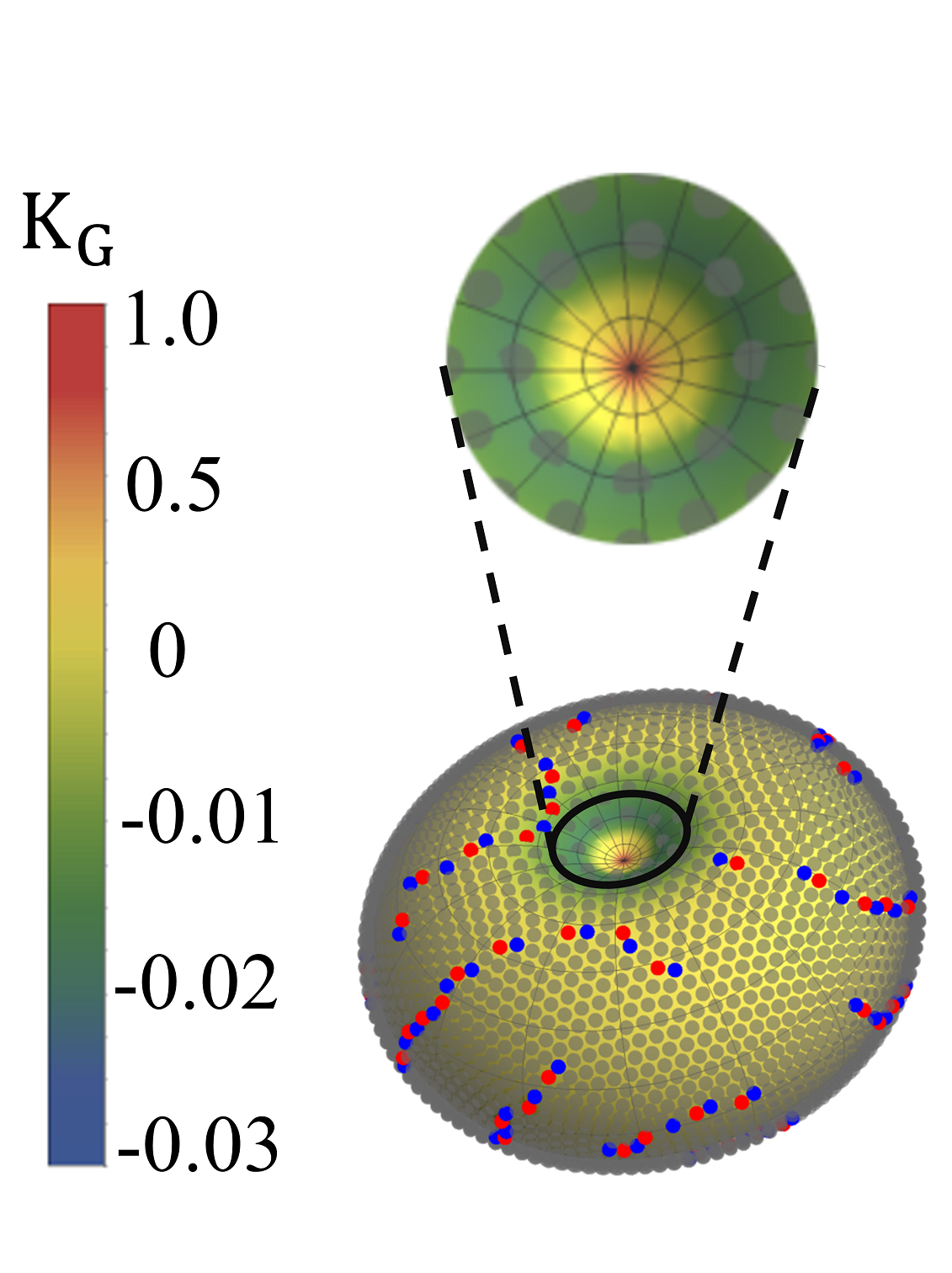}}
\subfigure[]{
\includegraphics[width=1.5in]{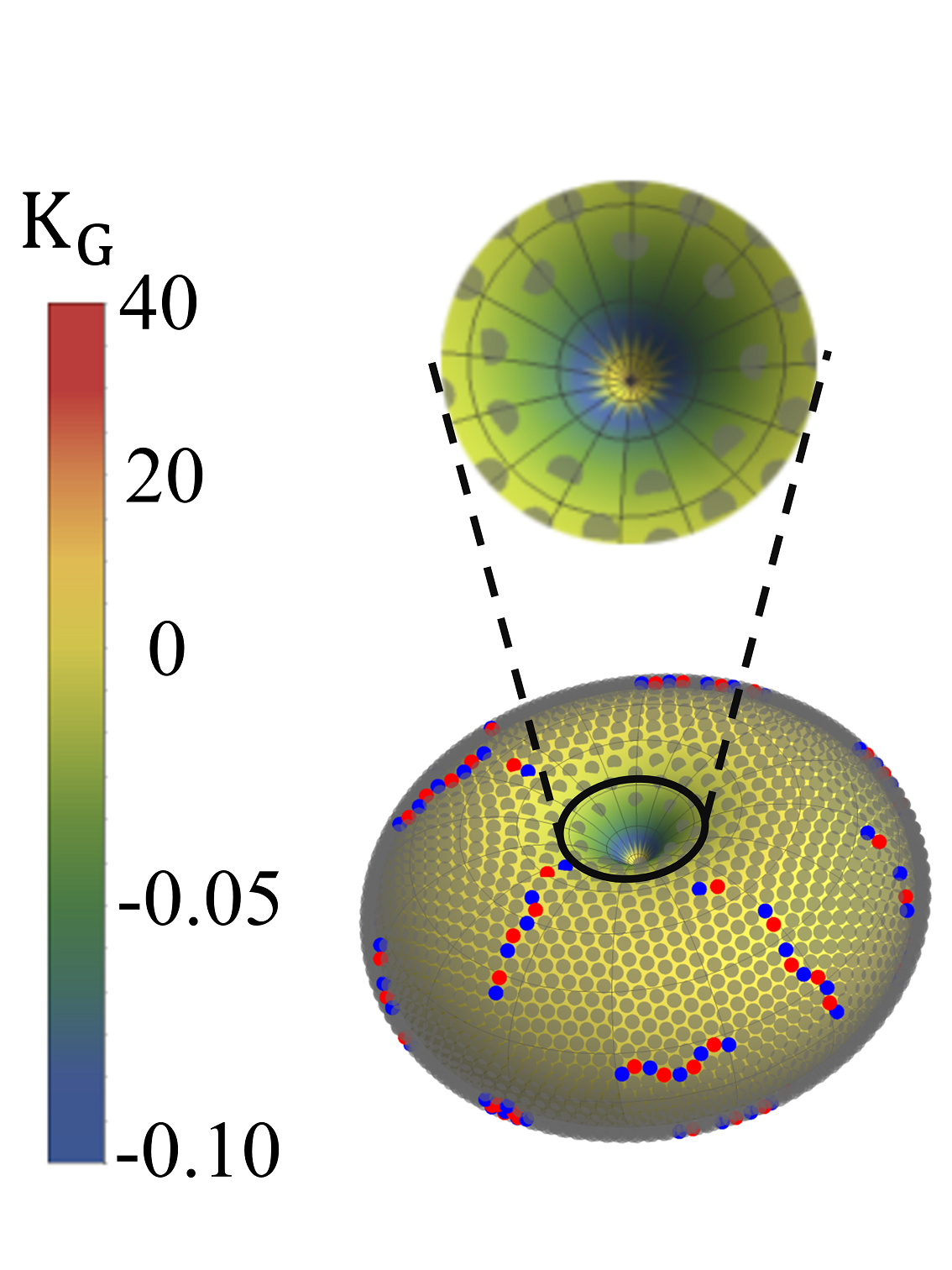}}
\vspace{-3mm}
\caption{Lowest-energy configurations on a pinched spheres with
    negative $\kappa$. (a) $\kappa=-0.1$, (b) $\kappa=-0.2$, (c) $\kappa=-0.3$, and (d) $\kappa=-0.4$. The five- and  seven-fold disclinations are colored in blue and red, respectively.
    $N=3000$.  }
\label{fig6}
\vspace{-3mm}
\end{figure}

\vspace{-2mm}

\subsection{Energetics analysis}

\begin{figure}[t]  
\centering 
\subfigure[]{
\includegraphics[width=1.63in]{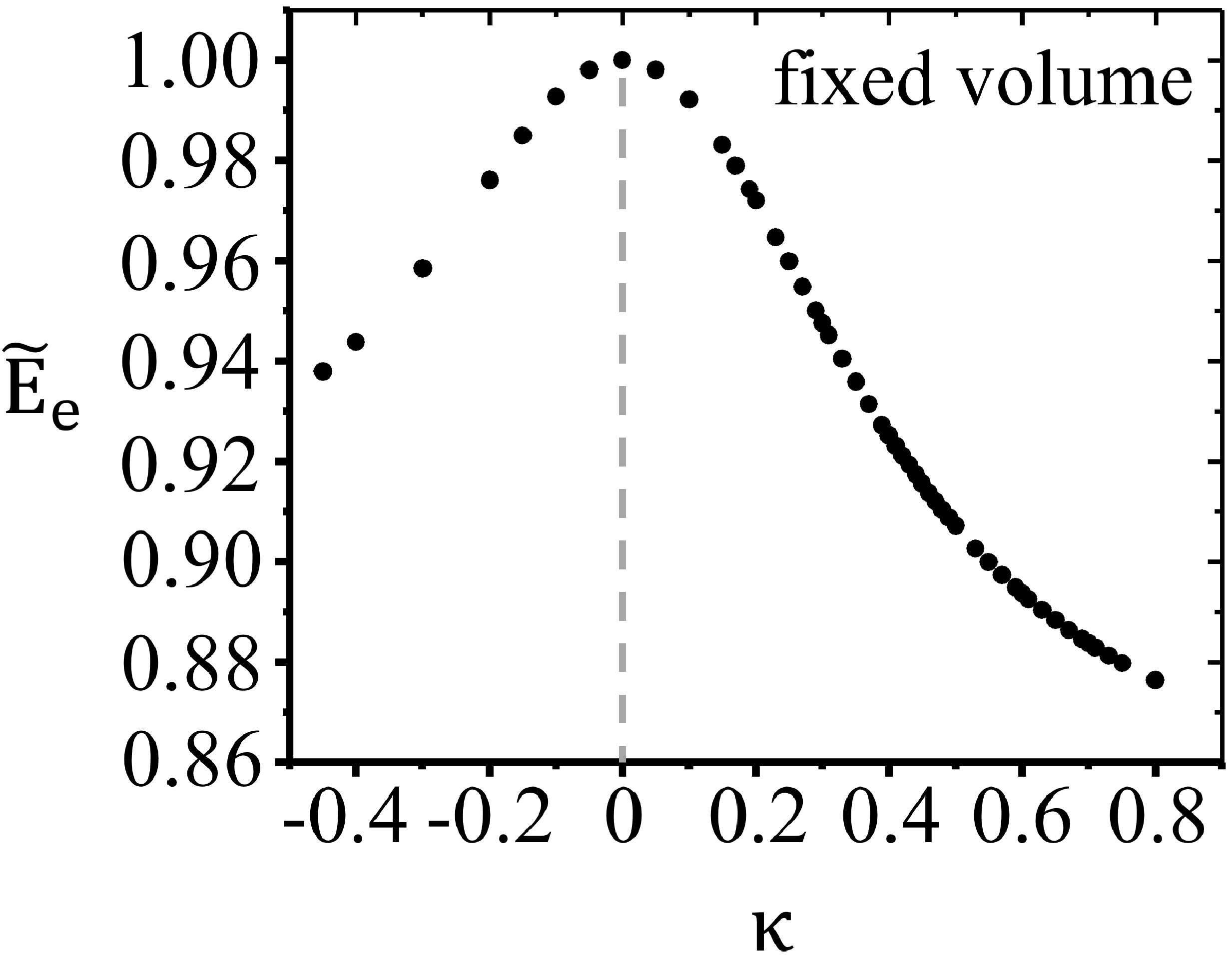}}
\hspace{-0.1in}
\subfigure[]{
\includegraphics[width=1.63in]{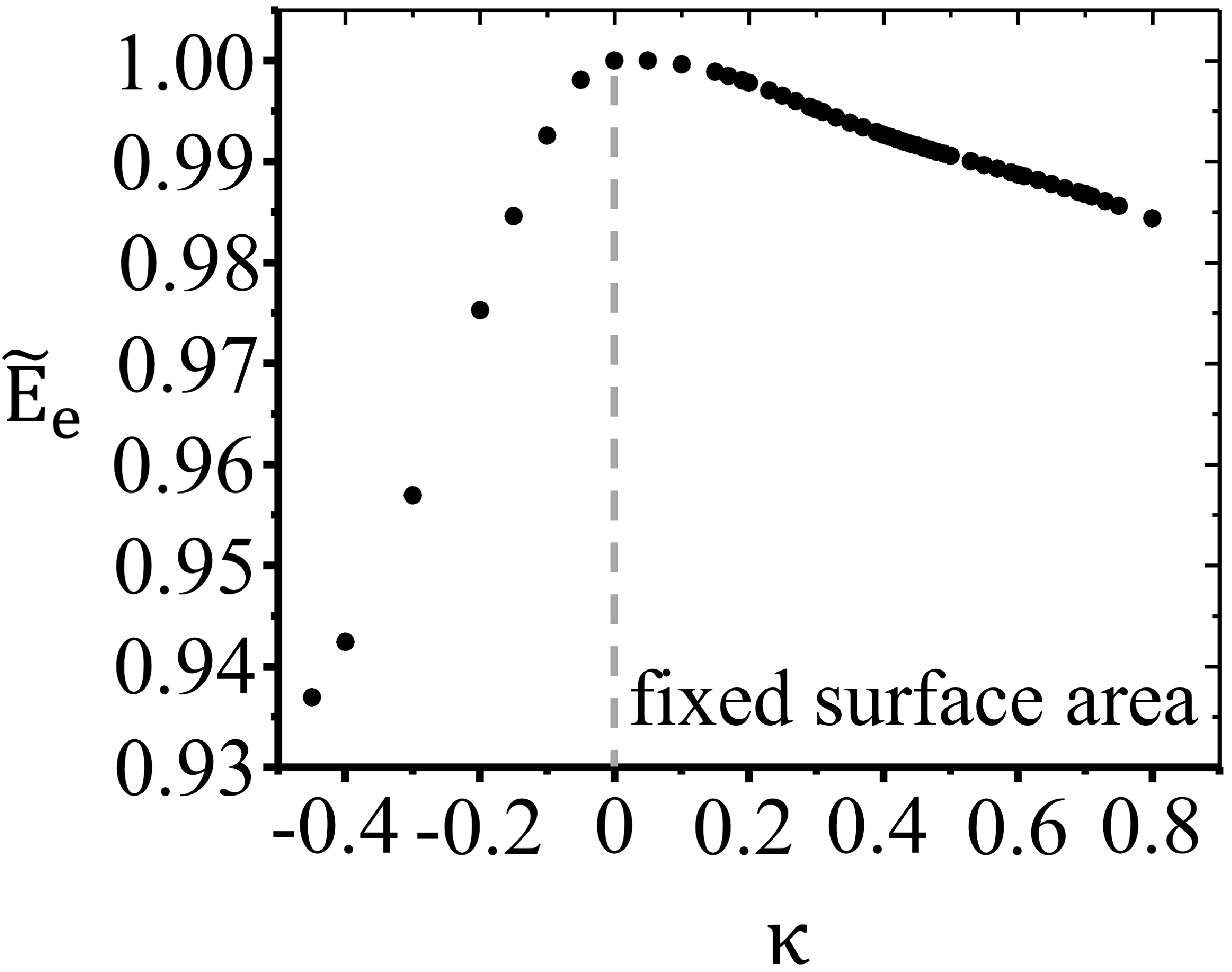}}
\caption{Electrostatic energy as a function of pinching parameter $\kappa$
  normalized by its value at $\kappa = 0$. (a) Volume is fixed, but surface
  area changes with $\kappa$; (b) Surface area is fixed, but volume changes with
  $\kappa$. $N=3000$.   }
\label{fig5}
\vspace{-3mm}
\end{figure}

So far, we have studied the packing of charged particles on frozen geometries.
Many surfaces in realistic systems are deformable. Here, we track the variation
of the total electrostatic energy as a function of the pinching parameter
$\kappa$, following two protocols: (1) fixed volume [see Fig.~\ref{fig5}(a)] and
(2) fixed surface area [see Fig.~\ref{fig5}(b)].  For both protocols, the total
electrostatic energy decreases monotonously as $\kappa$ deviate from zero.
This indicates, rather interestingly, that the surface will spontaneously deform
to a non-spherical shape as long as the non-electrostatic interactions, such as
surface tension and curvature energy, are sufficiently weak.  This shall be
discussed in a separate work. We also note that the reduction of the total
electrostatic energy indicates that pinching a charged sphere can increase its 
capacitance as an electric conductor, since $C=Q/U=Q^2/(2E)$, where $Q$ is the total
charge, $U$ is the electric potential, and $E$ is the electrostatic energy of
the conductor.

\vspace{-3mm}

\section{CONCLUSION}

In this work, the charged interface is deformed in a controllable fashion by
tuning the single parameter $\kappa$. This strategy has  its merit, but it
inevitably excludes the possibilities of other lowest-energy shapes. Rich
structures can be developed on a charged interface, such as the experimentally
observed singularity structures developed at the poles of a charged ellipsoidal
droplet~\cite{gomez1994charge, duft2003coulomb}, and the hierarchical
bucklings of an elastic charged ribbon~\cite{yao2016electrostatics}.
Furthermore, fluctuations of surface charges in real systems may modify the
bending rigidity and influence the stability of the charged
membranes~\cite{lau1998charge,kim2002effects, kim2003effects}. It is of great
interest to generalize our model by including these effects to approach a real
charged, deformable interface in the future work.

In summary, through the generalized Thomson problem, we study the combined
effects of long-range interaction and curvature in shaping the crystalline order
on the deformed sphere. As a key result in this work, we discover the depletion
zone structure, and reveal its physical origin as a finite size effect that
results from the geometry-regulated long-range interaction. These findings
provide further insights into the crystallography of charged particles on curved
surfaces. A future direction is to fully explore the shape space of a charged
interface of various topologies.  Notably, singularity structures may be
developed over a freely evolving charged interface under the long-range
interaction~\cite{gomez1994charge, duft2003coulomb}.

\section*{Acknowledgement}

Z.Y. and J.C. acknowledge support from NSFC Grant No.16Z103010253, the SJTU
startup fund under Grant No.WF220441904, and the award of the Chinese Thousand
Talents Program for Distinguished Young Scholars under Grants No.16Z127060004 and
No. 17Z127060032. X.X. acknowledges support from NSFC via Grant No. 11674217,
as well as Shanghai Municipal Education Commission and Shanghai Education
Development Foundation via the ``Shu Guang" project.


\begin{thebibliography}{49}
\expandafter\ifx\csname natexlab\endcsname\relax\def\natexlab#1{#1}\fi
\expandafter\ifx\csname bibnamefont\endcsname\relax
  \def\bibnamefont#1{#1}\fi
\expandafter\ifx\csname bibfnamefont\endcsname\relax
  \def\bibfnamefont#1{#1}\fi
\expandafter\ifx\csname citenamefont\endcsname\relax
  \def\citenamefont#1{#1}\fi
\expandafter\ifx\csname url\endcsname\relax
  \def\url#1{\texttt{#1}}\fi
\expandafter\ifx\csname urlprefix\endcsname\relax\def\urlprefix{URL }\fi
\providecommand{\bibinfo}[2]{#2}
\providecommand{\eprint}[2][]{\url{#2}}

\bibitem[{\citenamefont{Rayleigh}(1882)}]{rayleigh1882xx}
\bibinfo{author}{\bibfnamefont{L.}~\bibnamefont{Rayleigh}},
  \bibinfo{journal}{Philos. Mag.}
  \textbf{\bibinfo{volume}{14}}, \bibinfo{pages}{184} (\bibinfo{year}{1882}).

\bibitem[{\citenamefont{Duft et~al.}(2002)\citenamefont{Duft, Lebius, Huber,
  Guet, and Leisner}}]{duft2002shape}
\bibinfo{author}{\bibfnamefont{D.}~\bibnamefont{Duft}},
  \bibinfo{author}{\bibfnamefont{H.}~\bibnamefont{Lebius}},
  \bibinfo{author}{\bibfnamefont{B.}~\bibnamefont{Huber}},
  \bibinfo{author}{\bibfnamefont{C.}~\bibnamefont{Guet}}, \bibnamefont{and}
  \bibinfo{author}{\bibfnamefont{T.}~\bibnamefont{Leisner}},
  \bibinfo{journal}{Phys. Rev. Lett.} \textbf{\bibinfo{volume}{89}},
  \bibinfo{pages}{084503} (\bibinfo{year}{2002}).

\bibitem[{\citenamefont{Liu et~al.}(2015)\citenamefont{Liu, Wyss,
  Fernandez-Nieves, and Shum}}]{liu2015dynamics}
\bibinfo{author}{\bibfnamefont{Z.}~\bibnamefont{Liu}},
  \bibinfo{author}{\bibfnamefont{H.~M.} \bibnamefont{Wyss}},
  \bibinfo{author}{\bibfnamefont{A.}~\bibnamefont{Fernandez-Nieves}},
  \bibnamefont{and} \bibinfo{author}{\bibfnamefont{H.~C.} \bibnamefont{Shum}},
  \bibinfo{journal}{Phys. Fluids} \textbf{\bibinfo{volume}{27}},
  \bibinfo{pages}{082003} (\bibinfo{year}{2015}).

\bibitem[{\citenamefont{Huebner and Chu}(1971)}]{huebner1971instability}
\bibinfo{author}{\bibfnamefont{A.}~\bibnamefont{Huebner}} \bibnamefont{and}
  \bibinfo{author}{\bibfnamefont{H.}~\bibnamefont{Chu}}, \bibinfo{journal}{J.
  Fliud Mech.} \textbf{\bibinfo{volume}{49}}, \bibinfo{pages}{361}
  (\bibinfo{year}{1971}).

\bibitem[{\citenamefont{Guerrero et~al.}(2014)\citenamefont{Guerrero, Rivero,
  Gundabala, Perez-Saborid, and Fernandez-Nieves}}]{guerrero2014whipping}
\bibinfo{author}{\bibfnamefont{J.}~\bibnamefont{Guerrero}},
  \bibinfo{author}{\bibfnamefont{J.}~\bibnamefont{Rivero}},
  \bibinfo{author}{\bibfnamefont{V.~R.} \bibnamefont{Gundabala}},
  \bibinfo{author}{\bibfnamefont{M.}~\bibnamefont{Perez-Saborid}},
  \bibnamefont{and}
  \bibinfo{author}{\bibfnamefont{A.}~\bibnamefont{Fernandez-Nieves}},
  \bibinfo{journal}{P. Natl. Acad. Sci. U.S.A.} \textbf{\bibinfo{volume}{111}},
  \bibinfo{pages}{13763} (\bibinfo{year}{2014}).

\bibitem[{\citenamefont{Pairam and
  Fern{\'a}ndez-Nieves}(2009)}]{pairam2009generation}
\bibinfo{author}{\bibfnamefont{E.}~\bibnamefont{Pairam}} \bibnamefont{and}
  \bibinfo{author}{\bibfnamefont{A.}~\bibnamefont{Fern{\'a}ndez-Nieves}},
  \bibinfo{journal}{Phys. Rev. Lett.} \textbf{\bibinfo{volume}{102}},
  \bibinfo{pages}{234501} (\bibinfo{year}{2009}).

\bibitem[{\citenamefont{Fragkopoulos and
  Fern{\'a}ndez-Nieves}(2017)}]{fragkopoulos2017toroidal}
\bibinfo{author}{\bibfnamefont{A.~A.} \bibnamefont{Fragkopoulos}}
  \bibnamefont{and}
  \bibinfo{author}{\bibfnamefont{A.}~\bibnamefont{Fern{\'a}ndez-Nieves}},
  \bibinfo{journal}{Phys. Rev. E} \textbf{\bibinfo{volume}{95}},
  \bibinfo{pages}{033122} (\bibinfo{year}{2017}).

\bibitem[{\citenamefont{Aggeli et~al.}(2001)\citenamefont{Aggeli, Nyrkova,
  Bell, Harding, Carrick, McLeish, Semenov, and
  Boden}}]{aggeli2001hierarchical}
\bibinfo{author}{\bibfnamefont{A.}~\bibnamefont{Aggeli}},
  \bibinfo{author}{\bibfnamefont{I.~A.} \bibnamefont{Nyrkova}},
  \bibinfo{author}{\bibfnamefont{M.}~\bibnamefont{Bell}},
  \bibinfo{author}{\bibfnamefont{R.}~\bibnamefont{Harding}},
  \bibinfo{author}{\bibfnamefont{L.}~\bibnamefont{Carrick}},
  \bibinfo{author}{\bibfnamefont{T.~C.} \bibnamefont{McLeish}},
  \bibinfo{author}{\bibfnamefont{A.~N.} \bibnamefont{Semenov}},
  \bibnamefont{and} \bibinfo{author}{\bibfnamefont{N.}~\bibnamefont{Boden}},
  \bibinfo{journal}{P. Natl. Acad. Sci. U.S.A.} \textbf{\bibinfo{volume}{98}},
  \bibinfo{pages}{11857} (\bibinfo{year}{2001}).

\bibitem[{\citenamefont{Weingarten et~al.}(2014)\citenamefont{Weingarten,
  Kazantsev, Palmer, McClendon, Koltonow, Samuel, Kiebala, Wasielewski, and
  Stupp}}]{weingarten2014self}
\bibinfo{author}{\bibfnamefont{A.~S.} \bibnamefont{Weingarten}},
  \bibinfo{author}{\bibfnamefont{R.~V.} \bibnamefont{Kazantsev}},
  \bibinfo{author}{\bibfnamefont{L.~C.} \bibnamefont{Palmer}},
  \bibinfo{author}{\bibfnamefont{M.}~\bibnamefont{McClendon}},
  \bibinfo{author}{\bibfnamefont{A.~R.} \bibnamefont{Koltonow}},
  \bibinfo{author}{\bibfnamefont{A.~P.} \bibnamefont{Samuel}},
  \bibinfo{author}{\bibfnamefont{D.~J.} \bibnamefont{Kiebala}},
  \bibinfo{author}{\bibfnamefont{M.~R.} \bibnamefont{Wasielewski}},
  \bibnamefont{and} \bibinfo{author}{\bibfnamefont{S.~I.} \bibnamefont{Stupp}},
  \bibinfo{journal}{Nat. Chem.} \textbf{\bibinfo{volume}{6}},
  \bibinfo{pages}{964} (\bibinfo{year}{2014}).

\bibitem[{\citenamefont{Andelman}(1995)}]{andelman1995electrostatic}
\bibinfo{author}{\bibfnamefont{D.}~\bibnamefont{Andelman}},
  \bibinfo{journal}{Handb. Biol. Phys.}
  \textbf{\bibinfo{volume}{1}}, \bibinfo{pages}{603} (\bibinfo{year}{1995}).

\bibitem[{\citenamefont{Lipowsky and Sackmann}(1995)}]{lipowsky1995structure}
\bibinfo{author}{\bibfnamefont{R.}~\bibnamefont{Lipowsky}} \bibnamefont{and}
  \bibinfo{author}{\bibfnamefont{E.}~\bibnamefont{Sackmann}},
  \emph{\bibinfo{title}{Structure and Dynamics of Membranes}},
  vol.~\bibinfo{volume}{1} (\bibinfo{publisher}{Elsevier},
  \bibinfo{year}{1995}).

\bibitem[{\citenamefont{Genet et~al.}(2001)\citenamefont{Genet, Costalat, and
  Burger}}]{genet2001influence}
\bibinfo{author}{\bibfnamefont{S.}~\bibnamefont{Genet}},
  \bibinfo{author}{\bibfnamefont{R.}~\bibnamefont{Costalat}}, \bibnamefont{and}
  \bibinfo{author}{\bibfnamefont{J.}~\bibnamefont{Burger}},
  \bibinfo{journal}{Biophys. J.} \textbf{\bibinfo{volume}{81}},
  \bibinfo{pages}{2442} (\bibinfo{year}{2001}).

\bibitem[{\citenamefont{Kim and Sung}(2003)}]{kim2003effects}
\bibinfo{author}{\bibfnamefont{Y.~W.} \bibnamefont{Kim}} \bibnamefont{and}
  \bibinfo{author}{\bibfnamefont{W.}~\bibnamefont{Sung}},
  \bibinfo{journal}{Phys. Rev. Lett.} \textbf{\bibinfo{volume}{91}},
  \bibinfo{pages}{118101} (\bibinfo{year}{2003}).

\bibitem[{\citenamefont{Vernizzi and Olvera de~la
  Cruz}(2007)}]{vernizzi2007faceting}
\bibinfo{author}{\bibfnamefont{G.}~\bibnamefont{Vernizzi}} \bibnamefont{and}
  \bibinfo{author}{\bibfnamefont{M.}~\bibnamefont{Olvera de~la Cruz}},
  \bibinfo{journal}{P. Natl. Acad. Sci. U.S.A.} \textbf{\bibinfo{volume}{104}},
  \bibinfo{pages}{18382} (\bibinfo{year}{2007}).

\bibitem[{\citenamefont{Adamcik et~al.}(2010)\citenamefont{Adamcik, Jung,
  Flakowski, De~Los~Rios, Dietler, and Mezzenga}}]{adamcik2010understanding}
\bibinfo{author}{\bibfnamefont{J.}~\bibnamefont{Adamcik}},
  \bibinfo{author}{\bibfnamefont{J.-M.} \bibnamefont{Jung}},
  \bibinfo{author}{\bibfnamefont{J.}~\bibnamefont{Flakowski}},
  \bibinfo{author}{\bibfnamefont{P.}~\bibnamefont{De~Los~Rios}},
  \bibinfo{author}{\bibfnamefont{G.}~\bibnamefont{Dietler}}, \bibnamefont{and}
  \bibinfo{author}{\bibfnamefont{R.}~\bibnamefont{Mezzenga}},
  \bibinfo{journal}{Nat. Nanotechnol.} \textbf{\bibinfo{volume}{5}},
  \bibinfo{pages}{423} (\bibinfo{year}{2010}).

\bibitem[{\citenamefont{Yao and Olvera de~la
  Cruz}(2016)}]{yao2016electrostatics}
\bibinfo{author}{\bibfnamefont{Z.}~\bibnamefont{Yao}} \bibnamefont{and}
  \bibinfo{author}{\bibfnamefont{M.}~\bibnamefont{Olvera de~la Cruz}},
  \bibinfo{journal}{Phys. Rev. Lett.} \textbf{\bibinfo{volume}{116}},
  \bibinfo{pages}{148101} (\bibinfo{year}{2016}).

\bibitem[{\citenamefont{Winterhalter and
  Helfrich}(1988)}]{winterhalter1988effect}
\bibinfo{author}{\bibfnamefont{M.}~\bibnamefont{Winterhalter}}
  \bibnamefont{and} \bibinfo{author}{\bibfnamefont{W.}~\bibnamefont{Helfrich}},
  \bibinfo{journal}{J. Phys. Chem.} \textbf{\bibinfo{volume}{92}},
  \bibinfo{pages}{6865} (\bibinfo{year}{1988}).

\bibitem[{\citenamefont{Lau and Pincus}(1998)}]{lau1998charge}
\bibinfo{author}{\bibfnamefont{A.}~\bibnamefont{Lau}} \bibnamefont{and}
  \bibinfo{author}{\bibfnamefont{P.}~\bibnamefont{Pincus}},
  \bibinfo{journal}{Phys. Rev. Lett.} \textbf{\bibinfo{volume}{81}},
  \bibinfo{pages}{1338} (\bibinfo{year}{1998}).

\bibitem[{\citenamefont{Nguyen et~al.}(1999)\citenamefont{Nguyen, Rouzina, and
  Shklovskii}}]{nguyen1999negative}
\bibinfo{author}{\bibfnamefont{T.}~\bibnamefont{Nguyen}},
  \bibinfo{author}{\bibfnamefont{I.}~\bibnamefont{Rouzina}}, \bibnamefont{and}
  \bibinfo{author}{\bibfnamefont{B.}~\bibnamefont{Shklovskii}},
  \bibinfo{journal}{Phys. Rev. E} \textbf{\bibinfo{volume}{60}},
  \bibinfo{pages}{7032} (\bibinfo{year}{1999}).

\bibitem[{\citenamefont{Netz}(2001)}]{netz2001buckling}
\bibinfo{author}{\bibfnamefont{R.~R.} \bibnamefont{Netz}},
  \bibinfo{journal}{Phys. Rev. E} \textbf{\bibinfo{volume}{64}},
  \bibinfo{pages}{051401} (\bibinfo{year}{2001}).

\bibitem[{\citenamefont{Palmer et~al.}(2007)\citenamefont{Palmer, Velichko,
  Olvera~de La~Cruz, and Stupp}}]{palmer2007supramolecular}
\bibinfo{author}{\bibfnamefont{L.~C.} \bibnamefont{Palmer}},
  \bibinfo{author}{\bibfnamefont{Y.~S.} \bibnamefont{Velichko}},
  \bibinfo{author}{\bibfnamefont{M.}~\bibnamefont{Olvera~de La~Cruz}},
  \bibnamefont{and} \bibinfo{author}{\bibfnamefont{S.~I.} \bibnamefont{Stupp}},
  \bibinfo{journal}{Philos. T. Roy. Soc. A} \textbf{\bibinfo{volume}{365}},
  \bibinfo{pages}{1417} (\bibinfo{year}{2007}).

\bibitem[{\citenamefont{Moyer et~al.}(2012)\citenamefont{Moyer, Cui, and
  Stupp}}]{moyer2012tuning}
\bibinfo{author}{\bibfnamefont{T.~J.} \bibnamefont{Moyer}},
  \bibinfo{author}{\bibfnamefont{H.}~\bibnamefont{Cui}}, \bibnamefont{and}
  \bibinfo{author}{\bibfnamefont{S.~I.} \bibnamefont{Stupp}},
  \bibinfo{journal}{J. Phys. Chem. B} \textbf{\bibinfo{volume}{117}},
  \bibinfo{pages}{4604} (\bibinfo{year}{2012}).

\bibitem[{\citenamefont{Taylor}(1964)}]{taylor1964disintegration}
\bibinfo{author}{\bibfnamefont{G.}~\bibnamefont{Taylor}},
  \bibinfo{journal}{Proc. R. Soc. London, Ser. A}
  \textbf{\bibinfo{volume}{280}}, \bibinfo{pages}{383} (\bibinfo{year}{1964}).


\bibitem[{\citenamefont{Ziabicki}(1976)}]{ziabicki1976physical}
\bibinfo{author}{\bibfnamefont{A.} \bibnamefont{Ziabicki}},
  \emph{\bibinfo{title}{Fundamentals of Fibre Formation: The Science of
Fibre Spinning and Drawing}}
  (\bibinfo{publisher}{John Wiley \& Sons, New York}, \bibinfo{year}{1976}).




\bibitem[{\citenamefont{Bailey}(1988)}]{bailey1988electrostatic}
\bibinfo{author}{\bibfnamefont{A.~G.} \bibnamefont{Bailey}},
  \emph{\bibinfo{title}{Electrostatic Spraying of Liquids}}
  (\bibinfo{publisher}{John Wiley \& Sons, New York}, \bibinfo{year}{1988}).

\bibitem[{\citenamefont{Gomez and Tang}(1994)}]{gomez1994charge}
\bibinfo{author}{\bibfnamefont{A.}~\bibnamefont{Gomez}} \bibnamefont{and}
  \bibinfo{author}{\bibfnamefont{K.}~\bibnamefont{Tang}},
  \bibinfo{journal}{Phys. Fluids} \textbf{\bibinfo{volume}{6}},
  \bibinfo{pages}{404} (\bibinfo{year}{1994}).

\bibitem[{\citenamefont{Urbanski et~al.}(2017)\citenamefont{Urbanski, Reyes,
  Noh, Sharma, Geng, Jampani, and Lagerwall}}]{Urbanski2017}
\bibinfo{author}{\bibfnamefont{M.}~\bibnamefont{Urbanski}},
  \bibinfo{author}{\bibfnamefont{C.~G.} \bibnamefont{Reyes}},
  \bibinfo{author}{\bibfnamefont{J.}~\bibnamefont{Noh}},
  \bibinfo{author}{\bibfnamefont{A.}~\bibnamefont{Sharma}},
  \bibinfo{author}{\bibfnamefont{Y.}~\bibnamefont{Geng}},
  \bibinfo{author}{\bibfnamefont{V.~S.~R.} \bibnamefont{Jampani}},
  \bibnamefont{and} \bibinfo{author}{\bibfnamefont{J.~P.}
  \bibnamefont{Lagerwall}}, \bibinfo{journal}{J. Phys.-Condens. Mat.}
  \textbf{\bibinfo{volume}{29}}, \bibinfo{pages}{133003}
  (\bibinfo{year}{2017}).

\bibitem[{\citenamefont{Thomson}(1904)}]{thomson1904xxiv}
\bibinfo{author}{\bibfnamefont{J.~J.} \bibnamefont{Thomson}},
  \bibinfo{journal}{Philos. Mag.} \textbf{\bibinfo{volume}{7}},
  \bibinfo{pages}{237} (\bibinfo{year}{1904}).

\bibitem[{\citenamefont{Nelson}(2002)}]{nelson2002defects}
\bibinfo{author}{\bibfnamefont{D.~R.} \bibnamefont{Nelson}},
  \emph{\bibinfo{title}{Defects and Geometry in Condensed Matter Physics}}
  (\bibinfo{publisher}{Cambridge University Press, Cambridge, England},
  \bibinfo{year}{2002}).

\bibitem[{\citenamefont{De~Luca et~al.}(2005)\citenamefont{De~Luca, Rodrigues,
  and Levin}}]{de2005electromagnetic}
\bibinfo{author}{\bibfnamefont{J.}~\bibnamefont{De~Luca}},
  \bibinfo{author}{\bibfnamefont{S.~B.} \bibnamefont{Rodrigues}},
  \bibnamefont{and} \bibinfo{author}{\bibfnamefont{Y.}~\bibnamefont{Levin}},
  \bibinfo{journal}{Europhys. Lett.} \textbf{\bibinfo{volume}{71}},
  \bibinfo{pages}{84} (\bibinfo{year}{2005}).

\bibitem[{\citenamefont{Bowick and Giomi}(2009)}]{bowick2009two}
\bibinfo{author}{\bibfnamefont{M.~J.} \bibnamefont{Bowick}} \bibnamefont{and}
  \bibinfo{author}{\bibfnamefont{L.}~\bibnamefont{Giomi}},
  \bibinfo{journal}{Adv. Phys.} \textbf{\bibinfo{volume}{58}},
  \bibinfo{pages}{449} (\bibinfo{year}{2009}).

\bibitem[{\citenamefont{Koning and Vitelli}(2016)}]{koning2014crystals}
\bibinfo{author}{\bibfnamefont{V.}~\bibnamefont{Koning}} \bibnamefont{and}
  \bibinfo{author}{\bibfnamefont{V.}~\bibnamefont{Vitelli}},
  \emph{\bibinfo{title}{Crystals and Liquid Crystals Confined to Curved
  Geometries}} (\bibinfo{publisher}{John Wiley \& Sons, Hoboken},
  \bibinfo{year}{2016}).

\bibitem[{\citenamefont{Bausch et~al.}(2003)\citenamefont{Bausch, Bowick,
  Cacciuto, Dinsmore, Hsu, Nelson, Nikolaides, Travesset, and
  Weitz}}]{bausch2003grain}
\bibinfo{author}{\bibfnamefont{A.}~\bibnamefont{Bausch}},
  \bibinfo{author}{\bibfnamefont{M.}~\bibnamefont{Bowick}},
  \bibinfo{author}{\bibfnamefont{A.}~\bibnamefont{Cacciuto}},
  \bibinfo{author}{\bibfnamefont{A.}~\bibnamefont{Dinsmore}},
  \bibinfo{author}{\bibfnamefont{M.}~\bibnamefont{Hsu}},
  \bibinfo{author}{\bibfnamefont{D.}~\bibnamefont{Nelson}},
  \bibinfo{author}{\bibfnamefont{M.}~\bibnamefont{Nikolaides}},
  \bibinfo{author}{\bibfnamefont{A.}~\bibnamefont{Travesset}},
  \bibnamefont{and} \bibinfo{author}{\bibfnamefont{D.}~\bibnamefont{Weitz}},
  \bibinfo{journal}{Science} \textbf{\bibinfo{volume}{299}},
  \bibinfo{pages}{1716} (\bibinfo{year}{2003}).

\bibitem[{\citenamefont{Meng et~al.}(2014)\citenamefont{Meng, Paulose, Nelson,
  and Manoharan}}]{meng2014elastic}
\bibinfo{author}{\bibfnamefont{G.}~\bibnamefont{Meng}},
  \bibinfo{author}{\bibfnamefont{J.}~\bibnamefont{Paulose}},
  \bibinfo{author}{\bibfnamefont{D.~R.} \bibnamefont{Nelson}},
  \bibnamefont{and} \bibinfo{author}{\bibfnamefont{V.~N.}
  \bibnamefont{Manoharan}}, \bibinfo{journal}{Science}
  \textbf{\bibinfo{volume}{343}}, \bibinfo{pages}{634} (\bibinfo{year}{2014}).

\bibitem[{\citenamefont{Yao}(2017)}]{yao2017topological}
\bibinfo{author}{\bibfnamefont{Z.}~\bibnamefont{Yao}}, \bibinfo{journal}{Soft
  Matter} \textbf{\bibinfo{volume}{13}}, \bibinfo{pages}{5905}
  (\bibinfo{year}{2017}).

\bibitem[{\citenamefont{Altschuler et~al.}(1997)\citenamefont{Altschuler,
  Williams, Ratner, Tipton, Stong, Dowla, and Wooten}}]{altschuler1997possible}
\bibinfo{author}{\bibfnamefont{E.~L.} \bibnamefont{Altschuler}},
  \bibinfo{author}{\bibfnamefont{T.~J.} \bibnamefont{Williams}},
  \bibinfo{author}{\bibfnamefont{E.~R.} \bibnamefont{Ratner}},
  \bibinfo{author}{\bibfnamefont{R.}~\bibnamefont{Tipton}},
  \bibinfo{author}{\bibfnamefont{R.}~\bibnamefont{Stong}},
  \bibinfo{author}{\bibfnamefont{F.}~\bibnamefont{Dowla}}, \bibnamefont{and}
  \bibinfo{author}{\bibfnamefont{F.}~\bibnamefont{Wooten}},
  \bibinfo{journal}{Phys. Rev. Lett.} \textbf{\bibinfo{volume}{78}},
  \bibinfo{pages}{2681} (\bibinfo{year}{1997}).

\bibitem[{\citenamefont{Bowick et~al.}(2002)\citenamefont{Bowick, Cacciuto,
  Nelson, and Travesset}}]{bowick2002crystalline}
\bibinfo{author}{\bibfnamefont{M.}~\bibnamefont{Bowick}},
  \bibinfo{author}{\bibfnamefont{A.}~\bibnamefont{Cacciuto}},
  \bibinfo{author}{\bibfnamefont{D.~R.} \bibnamefont{Nelson}},
  \bibnamefont{and}
  \bibinfo{author}{\bibfnamefont{A.}~\bibnamefont{Travesset}},
  \bibinfo{journal}{Phys. Rev. Lett.} \textbf{\bibinfo{volume}{89}},
  \bibinfo{pages}{185502} (\bibinfo{year}{2002}).

\bibitem[{\citenamefont{Vitelli et~al.}(2006)\citenamefont{Vitelli, Lucks, and
  Nelson}}]{vitelli2006crystallography}
\bibinfo{author}{\bibfnamefont{V.}~\bibnamefont{Vitelli}},
  \bibinfo{author}{\bibfnamefont{J.~B.} \bibnamefont{Lucks}}, \bibnamefont{and}
  \bibinfo{author}{\bibfnamefont{D.~R.} \bibnamefont{Nelson}},
  \bibinfo{journal}{Proc. Natl. Acad. Sci. U.S.A.}
  \textbf{\bibinfo{volume}{103}}, \bibinfo{pages}{12323}
  (\bibinfo{year}{2006}).

\bibitem[{\citenamefont{Burke et~al.}(2015)\citenamefont{Burke, Mbanga, Wei,
  Spicer, and Atherton}}]{burke2015role}
\bibinfo{author}{\bibfnamefont{C.~J.} \bibnamefont{Burke}},
  \bibinfo{author}{\bibfnamefont{B.~L.} \bibnamefont{Mbanga}},
  \bibinfo{author}{\bibfnamefont{Z.}~\bibnamefont{Wei}},
  \bibinfo{author}{\bibfnamefont{P.~T.} \bibnamefont{Spicer}},
  \bibnamefont{and} \bibinfo{author}{\bibfnamefont{T.~J.}
  \bibnamefont{Atherton}}, \bibinfo{journal}{Soft Matter}
  \textbf{\bibinfo{volume}{11}}, \bibinfo{pages}{5872} (\bibinfo{year}{2015}).

\bibitem[{\citenamefont{Mehta et~al.}(2016)\citenamefont{Mehta, Chen, Chen,
  Kusumaatmaja, and Wales}}]{mehta2016kinetic}
\bibinfo{author}{\bibfnamefont{D.}~\bibnamefont{Mehta}},
  \bibinfo{author}{\bibfnamefont{J.}~\bibnamefont{Chen}},
  \bibinfo{author}{\bibfnamefont{D.~Z.} \bibnamefont{Chen}},
  \bibinfo{author}{\bibfnamefont{H.}~\bibnamefont{Kusumaatmaja}},
  \bibnamefont{and} \bibinfo{author}{\bibfnamefont{D.~J.} \bibnamefont{Wales}},
  \bibinfo{journal}{Phys. Rev. Lett.} \textbf{\bibinfo{volume}{117}},
  \bibinfo{pages}{028301} (\bibinfo{year}{2016}).

\bibitem[{\citenamefont{Yao}(2016)}]{yao2016dressed}
\bibinfo{author}{\bibfnamefont{Z.}~\bibnamefont{Yao}}, \bibinfo{journal}{Soft
  Matter} \textbf{\bibinfo{volume}{12}}, \bibinfo{pages}{7020}
  (\bibinfo{year}{2016}).

\bibitem[{\citenamefont{Irvine et~al.}(2010)\citenamefont{Irvine, Vitelli, and
  Chaikin}}]{irvine2010pleats}
\bibinfo{author}{\bibfnamefont{W.~T.} \bibnamefont{Irvine}},
  \bibinfo{author}{\bibfnamefont{V.}~\bibnamefont{Vitelli}}, \bibnamefont{and}
  \bibinfo{author}{\bibfnamefont{P.~M.} \bibnamefont{Chaikin}},
  \bibinfo{journal}{Nature (London)} \textbf{\bibinfo{volume}{468}},
  \bibinfo{pages}{947} (\bibinfo{year}{2010}).

\bibitem[{\citenamefont{Irvine et~al.}(2012)\citenamefont{Irvine, Bowick, and
  Chaikin}}]{irvine2012fractionalization}
\bibinfo{author}{\bibfnamefont{W.~T.} \bibnamefont{Irvine}},
  \bibinfo{author}{\bibfnamefont{M.~J.} \bibnamefont{Bowick}},
  \bibnamefont{and} \bibinfo{author}{\bibfnamefont{P.~M.}
  \bibnamefont{Chaikin}}, \bibinfo{journal}{Nat. Mater.}
  \textbf{\bibinfo{volume}{11}}, \bibinfo{pages}{948} (\bibinfo{year}{2012}).

\bibitem[{\citenamefont{Struik}(1988)}]{struik88a}
\bibinfo{author}{\bibfnamefont{D.}~\bibnamefont{Struik}},
  \emph{\bibinfo{title}{Lectures on Classical Differential Geometry}}
  (\bibinfo{publisher}{Dover Publications, New York}, \bibinfo{year}{1988}),
  \bibinfo{edition}{2nd} ed.

\bibitem[{\citenamefont{Greub et~al.}(1973)\citenamefont{Greub, Halperin, and
  Vanstone}}]{greub1973lie}
\bibinfo{author}{\bibfnamefont{W.}~\bibnamefont{Greub}},
  \bibinfo{author}{\bibfnamefont{S.}~\bibnamefont{Halperin}}, \bibnamefont{and}
  \bibinfo{author}{\bibfnamefont{R.}~\bibnamefont{Vanstone}},
  \emph{\bibinfo{title}{Lie Groups, Principal Bundles, and Characteristic
  Classes}} (\bibinfo{publisher}{Academic Press, New York and London},
  \bibinfo{year}{1973}).

\bibitem[{\citenamefont{Yao and Olvera de~la Cruz}(2013)}]{yao2013topological}
\bibinfo{author}{\bibfnamefont{Z.}~\bibnamefont{Yao}} \bibnamefont{and}
  \bibinfo{author}{\bibfnamefont{M.}~\bibnamefont{Olvera de~la Cruz}},
  \bibinfo{journal}{Phys. Rev. Lett.} \textbf{\bibinfo{volume}{111}},
  \bibinfo{pages}{115503} (\bibinfo{year}{2013}).

\bibitem[{\citenamefont{Mughal and Moore}(2007)}]{Mughal2007}
\bibinfo{author}{\bibfnamefont{A.}~\bibnamefont{Mughal}} \bibnamefont{and}
  \bibinfo{author}{\bibfnamefont{M.}~\bibnamefont{Moore}},
  \bibinfo{journal}{Phys. Rev. E} \textbf{\bibinfo{volume}{76}},
  \bibinfo{pages}{011606} (\bibinfo{year}{2007}).

\bibitem[{\citenamefont{Duft et~al.}(2003)\citenamefont{Duft, Achtzehn,
  M{\"u}ller, Huber, and Leisner}}]{duft2003coulomb}
\bibinfo{author}{\bibfnamefont{D.}~\bibnamefont{Duft}},
  \bibinfo{author}{\bibfnamefont{T.}~\bibnamefont{Achtzehn}},
  \bibinfo{author}{\bibfnamefont{R.}~\bibnamefont{M{\"u}ller}},
  \bibinfo{author}{\bibfnamefont{B.~A.} \bibnamefont{Huber}}, \bibnamefont{and}
  \bibinfo{author}{\bibfnamefont{T.}~\bibnamefont{Leisner}},
  \bibinfo{journal}{Nature (London)} \textbf{\bibinfo{volume}{421}},
  \bibinfo{pages}{128} (\bibinfo{year}{2003}).

\bibitem[{\citenamefont{Kim and Sung}(2002)}]{kim2002effects}
\bibinfo{author}{\bibfnamefont{Y.}~\bibnamefont{Kim}} \bibnamefont{and}
  \bibinfo{author}{\bibfnamefont{W.}~\bibnamefont{Sung}},
  \bibinfo{journal}{Europhys. Lett.} \textbf{\bibinfo{volume}{58}},
  \bibinfo{pages}{147} (\bibinfo{year}{2002}).

\end{thebibliography}
\end{document}